\documentclass[prb,twocolumn,superscriptaddress,english]{revtex4-2}

\usepackage{graphicx}
\usepackage{enumerate}
\usepackage[colorlinks=true,citecolor=blue]{hyperref}
\usepackage{textcomp}
\usepackage{amsmath}
\usepackage{amssymb}
\usepackage{pgf}
\usepackage[english]{babel}
\usepackage[normalem]{ulem}

\def\bra#1{\langle#1 |}
\def\ket#1{| #1\rangle}

\def\ud{\mathrm{d}}

\newcommand{\nep}{\textrm{e}}

\newcommand{\opc}[1]{{\hat{c}^{\phantom \dagger}}_{#1}}
\newcommand{\opcdag}[1]{{\hat{c}^{\dagger}}_{#1}}
\newcommand{\opd}[1]{{\hat{d}^{\phantom \dagger}}_{#1}}
\newcommand{\opddag}[1]{{\hat{d}^{\dagger}}_{#1}}

\newcommand{\opgamma}[1]{\hat{ \gamma}_{#1}}
\newcommand{\opgammadag}[1]{\hat{\gamma}^{\dagger}_{#1}}

\newcommand{\Ham}{\widehat{H}}
\newcommand{\Hleads}{\widehat{H}_{\rm leads}}
\newcommand{\Hsys}{\widehat{H}_{\rm sys}}
\newcommand{\Htunn}{\widehat{H}_{\rm tunn}}
\newcommand{\Hc}{\widehat{H}_{\rm c}}
\newcommand{\hc}{{\rm H.c.}}
\newcommand{\Vbias}{V_{b}}

\begin{document}

\title{Matrix product state simulations of quantum quenches and transport in Coulomb blockaded superconducting devices}

\author{Chia-Min Chung}
\affiliation{Niels Bohr International Academy and Center for Quantum Devices, Niels Bohr Institute, Copenhagen University, Universitetsparken 5, 2100 Copenhagen, Denmark}
\affiliation{Department of Physics, National Sun Yat-sen University, Kaohsiung 80424, Taiwan}
\affiliation{Center for Theoretical and Computational Physics, National Sun Yat-Sen University, Kaohsiung 80424, Taiwan}
\author{Matteo M. Wauters}
\affiliation{Niels Bohr International Academy and Center for Quantum Devices, Niels Bohr Institute, Copenhagen University, Universitetsparken 5, 2100 Copenhagen, Denmark}
\author{Michele Burrello}
\affiliation{Niels Bohr International Academy and Center for Quantum Devices, Niels Bohr Institute, Copenhagen University, Universitetsparken 5, 2100 Copenhagen, Denmark}

\begin{abstract}
Superconducting devices subject to strong charging energy interactions and Coulomb blockade are one of the key elements for the development of nanoelectronics and constitute common building blocks of quantum computation platforms and topological superconducting setups. The study of their transport properties is non-trivial and some of their non-perturbative aspects are hard to capture with the most ordinary techniques. 
Here we present a matrix product state approach to simulate the real-time dynamics of these systems. We propose a study of their transport based on the analysis of the currents after quantum quenches connecting such devices with external leads. Our method is based on the combination of a Wilson chain construction for the leads and a mean-field BCS description for the superconducting scatterers. In particular, we employ a quasiparticle energy eigenbasis which greatly reduces their entanglement growth and we introduce an auxiliary degree of freedom to encode the device total charge. 
This approach allows us to treat non-perturbatively both their charging energy and coupling with external electrodes.
We show that our construction is able to describe the Coulomb diamond structure of a superconducting dot with subgap states, including its sequential tunneling and cotunneling features. We also study the conductance zero-bias peaks caused by Majorana modes in a blockaded Kitaev chain, and compare our results with common Breit-Wigner predictions.
\end{abstract}

\maketitle

\section{Introduction}

The study of the transport properties of nanostructures is a pillar in the understanding of semiconducting and superconducting materials and in the engineering of devices for quantum technologies.
Quantum dots, single-electron transistors and superconducting Cooper pair boxes are fundamental building blocks of many of the envisioned devices for quantum information processing based on solid-state architectures, and the main diagnostic tools adopted in their experimental investigation are provided by tunneling spectroscopy.
These Coulomb blockaded elements play a crucial role in the design of novel platforms for nanoelectronics and hybrid heterostructures, as the ones adopted, for instance, for the fabrication of topological superconductors. It is therefore of the uttermost importance to develop suitable theoretical and numerical tools to simulate their dynamics and estimate the nonlinear conductance of such nanostructures characterized by strong charging energy effects. 

The most common strategies to model the transport across these interacting systems typically rely on master-equation approaches \cite{Beenakker1991} and perturbation theory over the coupling with the external leads (see, for example the review \cite{Aleiner2002}). Such techniques have been very fruitful in describing many blockaded devices but they often fail in capturing the emergence of non-perturbative phenomena, as the ones characterizing, for example, several impurity problems.

For strongly correlated quantum impurity systems, a plethora of perturbative renormalization group techniques have been developed since the Wilsonian formulation of the renormalization group and numerical renormalization group (NRG)~\cite{Wilson_RevModPhys75,Bulla_RevModPhys2008}. 
These methods are efficient to describe toy models based on a reduced number of degrees of freedom. They capture indeed the main universal features of the systems under investigation. However, they become computationally demanding when dealing with more realistic scenarios that include a larger number of degrees of freedom, and they suffer from limitations when dealing with nonequilibrium steady states \cite{PhysRevB.79.235336,Lotem2020}.
Therefore, the development of novel and complementary techniques to study the transport of complex many-body scatterers is a task of considerable importance in the modeling of nanodevices.

In this work, we present a non-perturbative strategy for the evaluation of the non-linear conductance of blockaded and superconducting quantum scatterers from a microscopic and out-of-equilibrium perspective. In particular, we will address systems with both strong charging energies and sizable couplings with the external leads, while accounting for the backaction of the scatterer onto the leads. 

In this respect, tensor networks \cite{SCHOLLWOCK201196,Silvi_SCIPOST2019} offer a very efficient set of tools to study the dynamics of quasi 1D interacting models, with the possibility of modelling interaction effects in a non-perturbative way, and limitations set instead by the growth of the entanglement of the system during its time evolution.

We will present an approach based on matrix product states (MPS) and the time dependent variational principle (TDVP) \cite{Haegeman_PRL2011,Haegeman2016} to estimate the conductance of blockaded devices through the real-time simulation of their dynamics. This allows us to evaluate the current as a function of time, to derive its transient behavior, and to estimate the transport features of these interacting systems in the stationary limit. We focus in particular on topological superconducting models, where the interplay between charging energy and superconductivity gives rise to the main signatures of Majorana modes observed so far in tunnel spectroscopy experiments \cite{Lutchyn_NatReview2018}.

Previous works have successfully applied time-dependent density matrix renormalization group (DMRG) to the study of conductance and noise in the interacting resonant level model out of equilibrium \cite{PhysRevB.70.121302,Schmitteckert2008,Schmitteckert2010,Schmitteckert2011,PhysRevB.79.235336,PhysRevB.82.205110,PhysRevB.73.195304,PhysRevLett.101.236801} and to the evaluation of the transport in simple nanostructures \cite{Schoenauer_PRR2019}.  Additionally, hybrid NRG-DMRG techniques have been adopted for simulating the quench dynamics of several quantum impurity models \cite{Guttge2013,Weichselbaum2018}. Very recently, the dynamics after a quench of the Anderson impurity models have also been studied through TDVP \cite{Rams2020,Kohn_PRB2021,Kohn_2022}.

Our approach is developed from similar techniques: we will consider systems composed by a Coulomb blockaded scatterer and two external leads and we will perform TDVP simulations of their dynamics after a quantum quench. Our algorithm adopts a Wilson chain description of the leads in terms of energy eigenstates inspired by NRG studies.  The scatterer is instead represented based on two main technical ingredients: (i) We adopt a BCS mean-field description and, in particular, we employ a (Bogoliubov) single-particle energy eigenstate basis to model the inner degrees of freedom of the scatterer in the MPS; this greatly reduces the entanglement growth during the evolution after the quench. (ii) We introduce an additional degree of freedom that keeps track of its charge dynamics; this allows us to capture the Coulomb charging energy of the system, obviating the violation of its particle number conservation caused by the BCS mean-field Hamiltonian \cite{keselman2019}. 
The combination of these two elements allows us to avoid the necessity of simulating an interacting number-conserving system (for instance the Richardson-Gaudin model~\cite{Dukelsky_RevModPhys2004}, see also the DMRG calculation of the spectra of superconducting models involving leads and quantum dots in Refs. \cite{saldana2021,zitko2021}), whose dynamics is typically difficult to simulate over sufficiently long time durations.

The quantum quench simulations we perform are reminiscent of the study of quenches in interacting one-dimensional models \cite{calabrese2016}, as, for example, the domain wall melting in the quantum XXZ chains (see, for instance, \cite{bertini2016,collura2018,biella2019,collura2020}). In integrable models, these real-time simulations are known to provide good electric and thermal conductance estimates, which typically match the Landauer B\"uttiker predictions obtained in a bosonization framework \cite{langmann2017}. Our calculations extend these results to general interacting models in which a Landauer B\"uttiker approach cannot be straightforwardly applied.

The rest of the paper is organized as follows:
In Sec.~\ref{sec:model} we describe the general structure of the blockaded systems we analyze. In Sec.~\ref{sec:quench} we discuss the general relation between quantum quench dynamics and transport properties in these systems.  In Sec.~\ref{sec:TTN} we outline the main features of our tensor network simulations and in Sec.~\ref{sec:benchmark} we present our results for two paradigmatic models of superconducting scatterers,  a generic p-wave superconducting quantum dot and the (blockaded) Kitaev chain, which provides a toy model for topological superconductors. For the former, our simulations provide estimates for the cotunneling conductance; for the latter, our non-perturbative approach predicts zero-bias Majorana peaks with a reduced amplitude and an enhanced width with respect to standard scattering matrix approaches. Finally we illustrate our conclusions in Sec.~\ref{sec:conclusion}. Additional details on the symmetries of our MPS construction and the rate equation estimates are presented in the appendices.

\section{Model}\label{sec:model}
In this work, we analyze transport problems in Coulomb blockaded devices connected with external one-dimensional metallic leads.
To illustrate our method, we focus on systems of spinless fermions and only two leads; however the generalization to spinful models and multiple leads is straightforward.
\begin{figure}
\begin{center}
\includegraphics[width=\columnwidth]{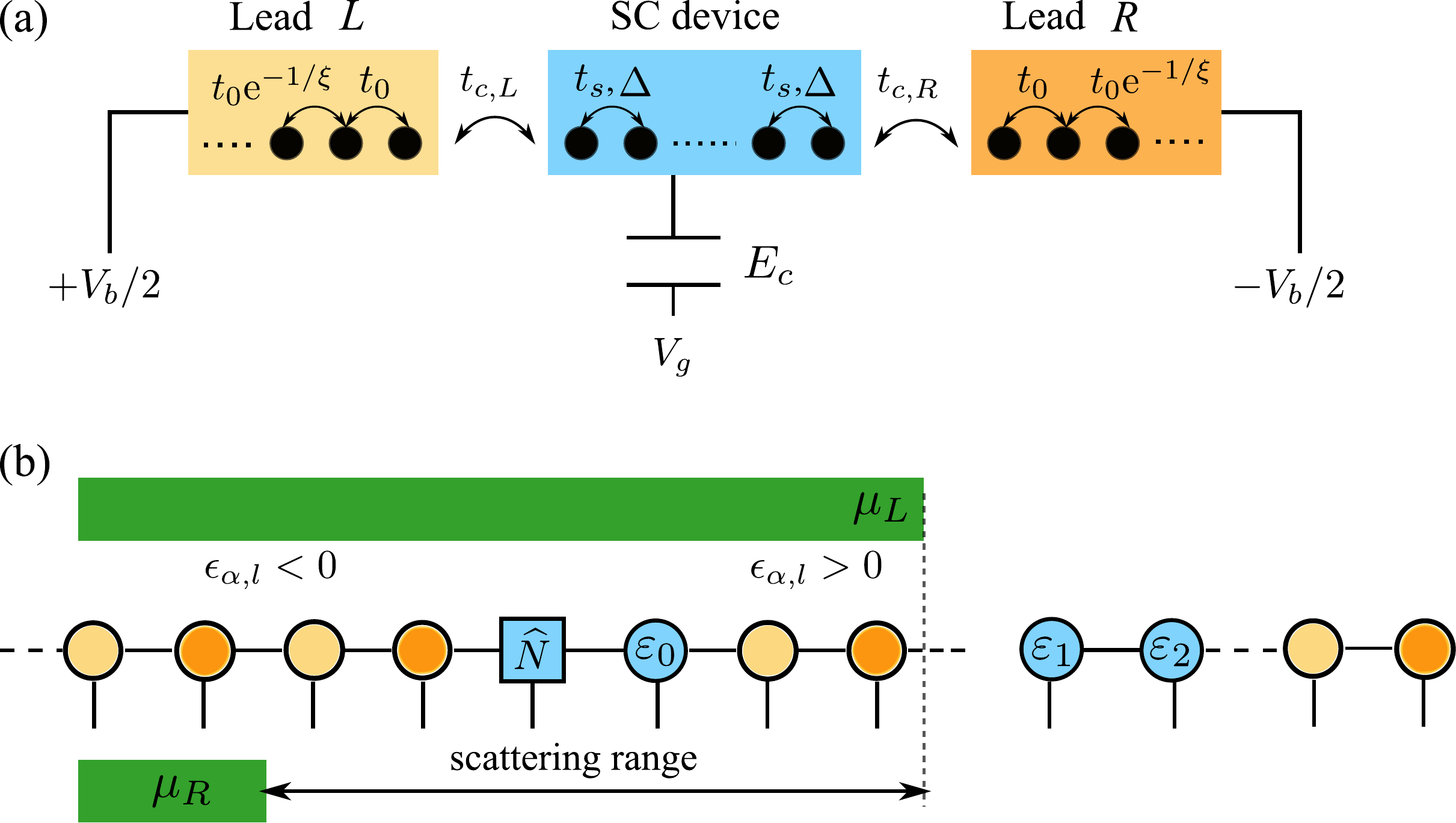}
\caption{(a) Sketch of the system we model with matrix product states: a floating superconducting (SC) island is connected to two normal leads, where we measure a single electron current. The gate voltage $V_g$ is used to tune the induced charge on the superconductor $n_g=\frac{e V_g}{2E_c}$.
(b) Matrix product state representation of the system. Each MPS site corresponds to a single-particle energy level of either the leads or the SC device (for $t_{c,\alpha}=0$) 
Yellow and orange circles represent the single-particle eigenstates of the left and right leads, respectively, ordered by their energies $\epsilon_{\alpha,l}$, whereas blue circles are the quasiparticle states of the superconductor, with energies $\varepsilon_n$. The square MPS site is the bosonic auxiliary site encoding the charge of the device.   
The leads energy levels are filled up to the corresponding chemical potential $\mu_{L/R}$; the states in the interval $\Vbias=\mu_L-\mu_R$ are those mostly involved in the transport process.}\label{fig:sketch}
\end{center}
\end{figure}

The general structure of the Hamiltonian we study is 
\begin{equation} \label{Hamtot}
\Ham_T = \Hleads + \Hsys + \Hc + \Htunn  \ ,
\end{equation}
where $\Hleads$ describes the two metallic leads; $\Hsys$ defines the scatterer device whose transport properties are under scrutiny, and it may include interactions and a mean-field superconducting BCS pairing; $\Hc$ determines its charging energy, which is the specific interaction responsible for the Coulomb blockaded regime; finally, $\Htunn$ represents the tunneling Hamiltonian between the device and the leads (see the schematic representation in Fig. \ref{fig:sketch}).

The paradigmatic system we consider is a one dimensional p-wave topological superconductor, hence we describe the scatterer as a Kitaev chain \cite{Kitaev_2001} with open boundaries.
Its charging energy is given by the electrostatic repulsion arising from the finite capacitance of the floating --i.e. not grounded-- device. The scatterer energy is thus defined by the following Hamiltonian contributions:
\begin{align}
\Hsys &= \sum_{j=1}^\mathcal{M} \left[ -t_s \opddag{j+1} \opd{j} + \Delta \opddag{j+1} \opddag{j} + \hc \right] -\mu_s \opddag{j}\opd{j} \,, \label{kitaevham} \\
\Hc &= E_c (\widehat{N}- n_g)^2 \, , \label{Hc}
\end{align}
where $t_s$ is the nearest-neighbors hopping amplitude, $\Delta$ the p-wave supeconducting pairing and $\mu_s$ the chemical potential. 
$E_c = \frac{e^2}{2C}$ is the energy associated to the addition of a single charge $e$ to the system, which here we consider to have an effective capacitance $C$. 
The electrostatic energy then depends on the difference between the charge of the scatterer $\widehat{N}$ and the charge $n_g$, typically induced in the experiments by a tunable voltage gate. Importantly, the Kitaev Hamiltonian does not conserve the particle number: superconductivity is indeed included in a BCS mean-field approximation. The device charge $\widehat{N}$, instead, is a conserved quantity if the scatterer is isolated and it accounts for the charge of both the Cooper pair condensate and the quasiparticle excitations. 
In particular, the total charge $\widehat{N}$ is different from the operator $\sum_j \opddag{j} \opd{j}$, which takes into account only the quasiparticle contribution to the charge; however, they share the same fermionic parity. 
We emphasize that the Hamiltonian $\Hsys$ can be easily generalized to different lattice models; in particular, short-range interactions can also be included with additional but affordable computational cost.
Hereafter we will label with $\mathcal{M}$ the number of sites involved in $\Hsys$.

The leads are described by simple one-dimensional nearest-neighbor hopping Hamiltonians 
\begin{equation} \label{Hleads}
\Hleads = \sum_{\alpha=L,R} \sum_{l=1}^\mathcal{L} \left[ -t^{\alpha}_{0l} \opcdag{\alpha,l+1}\opc{\alpha,l} +\hc \right] - \mu_\alpha \opcdag{\alpha,l}\opc{\alpha,l} \ , 
\end{equation}
where $\alpha = L,R$ labels the two different leads.
The sites are ordered such that in both leads we start counting from the contact with the scatterer.
Notice also that the hopping amplitude can be site dependent: specifically, we choose an exponential decay of the form 
\begin{equation} \label{Wilsonchain}
 t_{0l}^\alpha = t_0 \nep^{-(l-1)/\xi} \ ,
\end{equation}
with a decay length $\xi < \mathcal{L}$, building what is known in the literature as Wilson chain~\cite{Wilson_RevModPhys75,Mitchell_PRB2004,DaSilva_PRB2008}.
Physically, it corresponds to a logarithmic discretization of the original continuous Hamitlonian, where the lattice spacing increases as we move farther away from the scatterer. 
This choice provides two main advantages, which will become clearer in the following section: 
first, there are more eigenvalues with energies close to the Fermi level, which increases our resolution close to zero voltage bias in the quench protocols;
second, the exponentially decaying hopping prevents the current to reach the end of the leads, acting as a sort of effective sink. 
This improves the convergence in time towards the intermediate quasi-steady-state we are interested in to compute transport properties. 
In the rest of the paper, we consider the two leads as equivalent under inversion symmetry, so we can drop the $\alpha$ index in the hopping amplitude, and we set the zero of the energy at the Fermi level corresponding to half-filled leads.

Finally, the tunneling between leads and system is described by
\begin{equation}
\Htunn = -t_{c,L} \left(\opddag{1}\opc{L,1} +\hc \right) -t_{c,R} \left(\opddag{\mathcal{M}}\opc{R,1} +\hc \right) \ .
\end{equation}
In the following, we will also consider $t_{c,L} =t_{c,R} = t_c$ for the sake of simplicity.

\section{Transport properties from quantum quenches}\label{sec:quench}
Our main goal is to compute the conductance of a given device from its nonequilibrium dynamics, without relying on perturbative approaches.
To this purpose, we simulate the time evolution of the system after a quantum quench and extract the conductance from the emerging quasi-steady-state.

\begin{figure*}[t]
  \begin{center}
    \includegraphics[width=18cm]{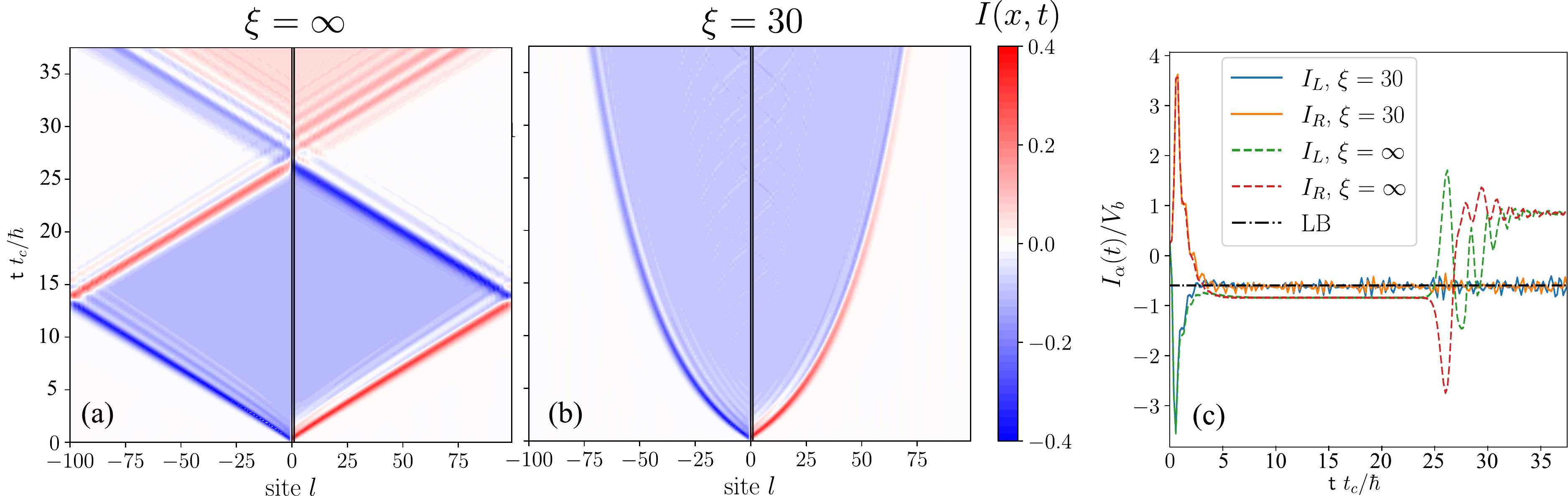} 
    \caption{Exact evolution of a non-interacting impurity model. (a) Space-time profile of the current, with uniform hopping in the leads. The vertical black lines mark the position of the impurity site. 
    (b) Space-time profile of the current when the hopping in the leads decays exponentially with a characteristic length $\xi=30$. 
    (c) Comparison between the time evolution of the current at the left and right edge of the scattering region, for uniform leads ($\xi = \infty$, dashed lines) and Wilson chains ($\xi =30$, solid lines).
    The horizontal dot-dashed line indicates the Landauer-B\"uttiker prediction.
    The energy of the impurity site is $\varepsilon_{\rm imp}=0.1t_0$ and the bias is chosen to be in resonance, hence $V_{b} = \varepsilon_{\rm imp}$. The other parameters are $\mathcal{L}=100$ and $t_c =0.25 t_0$.}\label{fig:quench_exact}
  \end{center}
\end{figure*}

In the thermodynamic limit and for noninteracting models,  the system relaxes toward a nonequilibrium steady-state with a current flow corresponding to the prediction of the Landauer-B\"uttiker (LB) formula \cite{Ljubotina_SciPost2019} after a quench in which leads with different densities are suddenly connected at time ${\sf t}=0$ ($t_c({\sf t}) \propto \Theta({\sf t})$, with $\Theta$ being the Heaviside step function).
In finite systems, instead, the LB regime appears only as a transient \cite{Chien2014} before the current is reflected back from the edges (see Fig. \ref{fig:quench_exact} for a qualitative example) and eventually a trivial steady-state is reached.
An estimate of the conductance of the system must rely on this transient non-equilibrium quasi steady state (NEQSS) which is, therefore, the object of our investigation. 
When the lead Hamiltonian $\Hleads$ includes only local terms, the post-quench time evolution obeys a Lieb-Robinson bound \cite{liebrobinson,bonnes2014}; the central scatterer region is not affected by the finite size effects of the simulation until the quasiparticles excited by the quench in the central region reach the boundaries and come back. Therefore we can consider the NEQSS as a faithful representation of what happens in the thermodynamic limit (see, for instance, \cite{Ljubotina_SciPost2019,bertini2016,Viti_2016,Essler_2016}). 

To bring the system out of equilibrium, we can consider two different quench protocols \cite{Chien2014}:
\begin{itemize}
\item $\mathbf{\mu}$-{\bf quench}: the system is initially prepared in the ground state of $\Ham_T$ where both leads have the same chemical potential. At $t=0$, a voltage bias $\Vbias = \mu_L -\mu_R$ is introduced and the system evolves accordingly to the new Hamiltonian $\Ham_T(\Vbias)$.

\item {\bf density quench}: the device and the leads are initially decoupled and the system is prepared in the ground state of $\Hsys + \Hleads$, where a voltage bias $\Vbias$ is used to induce a density difference between the two leads. At ${\sf t}=0$, the bias is turned off and the system evolves with $\Ham_T$. 
\end{itemize}

After the quench, me measure the current in the leads, following the definition
\begin{equation} \label{currentop}
I_{\alpha,l}({\sf t}) = 2\pi i t^\alpha_{0l}  \bra{\Psi({\sf t})} \opcdag{\alpha,l+1}\opc{\alpha,l} - \opcdag{\alpha,l}\opc{\alpha,l+1} \ket{\Psi({\sf t})}  \ ,
\end{equation}
where $\ket{\Psi({\sf t})}$ is the time-evolved many-body wavefunction. 

Although for small biases the two quench protocols give consistent results, they are not exactly equivalent.
This can be easily seen if we consider a non-interacting case and the related structure of its scattering matrix in one dimension.
In the $\mu$-quench, the leads have different chemical potential, meaning that an energy step is added on top of the scattering matrix of the device. Hence, momentum is not conserved when a particle is transmitted across the device.
In the density quench, instead, far from the device the two leads have the same chemical potential and the momentum is a good quantum number for the scattering process.

In our simulation, we have verified that the density quench protocol gives usually better results for the estimates of the conductance, both in terms of a shorter relaxation time before reaching the quasi-steady-state and a more stable current profile.
Thus, unless otherwise stated, all the results we present are obtained with the density quench protocol. Concerning the simulation of the time-evolution of the system, however, both quench protocols can be 
implemented with analogous accuracy.

To illustrate the main physical properties and limitations of studying transport properties through quantum quenches, we discuss next the performance of the method on a noninteracting impurity model.
First, let us consider the situation where the leads have a uniform hopping amplitude, corresponding to an infinite decay length $\xi \to \infty$.
The single site impurity with energy $\varepsilon_{\rm imp}$ is described by the Hamiltonian $\Hsys=\varepsilon_{\rm imp} \opddag{} \opd{}$.

In Fig.~\ref{fig:quench_exact}, we summarize the main characteristics of the quench protocol.
The system is initialized with a density imbalance between the left and the right leads, which induces a current flow in the chain. 
At first, the current involves only the sites immediately adjacent to the impurity. The initial discontinuity in the density profile splits into two fronts counterpropagating along the two leads  with constant speed in the whole system.
This is clearly shown in Fig.~\ref{fig:quench_exact}(a), where we plot the evolution of current density $I({\sf t})$ in time and space. 
In particular we emphasize than inside this broadening ``cone'' a NEQSS emerges, carrying a steady current consistent with the LB result, as seen in Fig.~\ref{fig:quench_exact}(c).

However, the finite length of the chain has two consequences:
the first, and more obvious, is that the current is reflected back from the edges of the leads and the NEQSS changes when this signal reaches again the edges of the scatterer.
In general, different NEQSS exist inside each of these rhomboids confined by the propagating signal, as suggested by the different plateaus displayed by the data corresponding to uniform leads $(\xi=\infty$, dashed lines) in Fig.~\ref{fig:quench_exact}(c).
Eventually, the system will reach a trivial steady-state with zero current.

The second consequence is the limit on the energy resolution due to the finite level spacing in the leads which is of the order $2t_0/\mathcal{L}$. 
This becomes particularly important when computing the current from a small voltage bias, because in the initial state the particle number in the two leads will differ only by a few units.
As a consequence, the current develops finite size corrections which deviate from the exact LB formula, again visible from Fig.~\ref{fig:quench_exact}(c). 

Both these problems are partially cured by choosing a finite decay length for the leads' hopping amplitude, as reported in panels (b) and (c) of Fig.~\ref{fig:quench_exact}.
Indeed, now the reflection from the edges of the system is strongly suppressed and quasi-stationary state survives for longer time.
Also the finite size corrections with respect to the LB formula are reduced, at the cost of a noisier current signal. 
However, this noise is easily eliminated by averaging the current in time, after the plateau is reached.

Physically, the Wilson chain construction in Eqs. (\ref{Hleads},\ref{Wilsonchain}) can be thought as a real space renormalization applied to the leads~\cite{Wilson_RevModPhys75,Mitchell_PRB2004}, where the further we are from the contact with the scatterer, the more chain sites are merged together. 
By comparing panels (a) and (b) in Fig. \ref{fig:quench_exact}, it is easy to realize that the introduction of the exponential dumping of the tunneling amplitude $t_0$ amounts to a compression of the spreading correlation cone.
While in the uniform chain the signal travels with constant Fermi speed $v_f=2t_0$, leading to a space-time profile ${\sf t} = \frac{l-1}{2t_0}$, in the Wilson chain the speed is exponentially damped. 
This results in the space-time profile ${\sf t} = \frac{\xi}{2 t_0}\left(\nep^{(l-1)/\xi}-1 \right)$.
In this way, we can simulate effectively larger system sizes, thus reducing the finite size effects.
This construction modifies also the spectrum of the leads: the eigenvalue spacing is denser at low energies, thus allowing a finer resolution at small voltage biases.

The main difficulty of this approach is the choice of a proper decay length; if $\xi$ is too large, its effect is negligible and if it is too small the current is reflected back from the effective edge created by the vanishing hopping amplitude.
This trade-off also depends on the strength of the bias, since larger values of $\Vbias$ also require larger decay length.
In our MPS simulations, we adjust heuristically the value of $\xi$ by choosing a value that does not introduce a nonphysical reflection in the current before the lead edges.

For the rest of the paper, unless otherwise specified, the leads will be described by two chains of $\mathcal{L}=100$ sites, with a bare hopping amplitude $t_0=1$ that sets all other energy scales in the model and a decaying length $\xi=40$.
The latter has been chosen in such a way that all the data we present do not display an artificial reflection of the current from the edges and the signal converges fast enough enabling us to average it over a sufficiently large time interval.

\section{Matrix product state implementation}\label{sec:TTN}

We implement the quench dynamics of the systems by using matrix product state (MPS) techniques~\cite{SCHOLLWOCK201196}. As all tensor network techniques, the underlying approximation restricts the maximal entanglement allowed in the simulated state at any time. In particular, we adopt a maximum bond dimension $\chi$ which varies as a function of time and is not uniform in the MPS construction.
In the following numerical simulations, the maximum bond dimension is limited by $\chi \lesssim 2500$, which implies that the simulation of the time evolution of the system is reliable until the correlations between its partitions are below a suitable threshold, which can be estimated by the maximum value of the entanglement entropy $S_{\rm max} = \log_2 \chi$.

Concerning the quench protocol, we will focus on the density quench presented in Sec.~\ref{sec:quench}: the scatterer is initially decoupled and the voltage bias $V_b$ is symmetrically applied ($\mu_L = -\mu_R = V_b/2$) to induce different electron densities in the two leads before the quench, corresponding to their Fermi-Dirac distribution at zero temperature.
The initial state is thus composed by the product of the independent ground states of the leads and the scatterer. 
The ground state in the non-interacting leads is simply a free-electron state; however, interacting leads can be considered as well and they can be initialized through density matrix renormalization group (DMRG) \cite{White1992,White1993} calculations.
DMRG can be adopted to initialize the scatterer in its ground state as well. 
At time ${\sf t}>0$, the system is quenched to a Hamiltonian with no bias, while the scatterer and the leads are coupled by the tunneling interaction $\Htunn$.

Our aim is to tackle superconducting and blockaded devices, with the possibility of describing the out-of-equilibrium physics of setups with large voltage bias and sizable tunneling interactions between the leads and the scatterer. In the following, we discuss in detail the construction of our computational basis, its consequences and the introduction of the charging energy in a number non-conserving model.

The method is implemented by using ITensor library \cite{itensor}. The source code can be found in the repository: \url{https://github.com/chiaminchung/QuenchTransport}.

\subsection{Energy basis}\label{ssec:ebasis}

A well-known challenge for the simulation of the dynamics of out-of-equilibrium many-body systems with tensor network techniques is the growth of their entanglement in time, which typically prevents the possibility of reaching an accurate description of the system dynamics for long times \cite{Schuch2008}. This is clearly a major limitation, since the estimate of the conductance of a given scatterer is based on the NEQSS behavior of the current. When considering an MPS based on a real space basis, the entanglement growth is clearly captured by the entanglement entropy $S$ associated to any partition that separates the degrees of freedom in different leads. Its linear growth in time is easily understood due to the fact that each particle coming from the leads is partially reflected and partially transmitted in a coherent way by the scatterer, thus generating entangled particle-hole pairs \cite{Rams2020,Kohn_PRB2021}. This results indeed in a linear increment of $S\propto \Vbias {\sf t}$ in time and voltage bias, which is consistent with the Lieb-Robinson picture of the dynamics of the system after a quench: the NEQSS appearing inside its space-time cone is progressively constituted by more and more particle-hole pairs, and their number increases linearly in time.

To mitigate the restrictions imposed to tensor network dynamics simulations by this growth of the entanglement, Rams and Zwolak proposed to adopt a different basis \cite{Rams2020}: when the leads are modelled based on an ordered single-particle energy eigenbasis, only states lying in an intermediate energy window defined by the voltage bias are affected in a major way by the time-evolution. Ref. \cite{Rams2020} shows indeed that by modeling a system with a mixed eigenbasis - an energy eigenbasis for the leads and a spacial basis for the scatterer - the entanglement entropy typically displays only a logarithmic growth in time, thus allowing for simulations of a considerably longer time evolution (see also \cite{Zhuoran2017}).  

In the following, we adopt this strategy and we use a single-particle eigenbasis of $\Hleads$ for the leads, with MPS sites ordered by increasing energy. The Wilson chain approach allows us to get a higher resolution in energy close to the Fermi energy, thus in the energy window in which the dynamics has major effects. Concerning the scatterer sites, we may adopt different bases depending on the physical system. 
The simplest choice is a real-space basis, in which the scatterer sites are kept together, ordered by their positions and located in proximity of the zero-energy single-particle states of the leads. This choice is convenient in the presence of local interactions within the scatterer Hamiltonian $\Hsys$. 

When the only interaction is provided by the charging energy $\Hc$, instead, it is convenient to apply a quasiparticle energy eigenbasis also for the scatterer. This will be our choice in the following, since we will focus on superconducting systems with a quadratic $\Hsys$. In the resulting MPS construction, the single-particle eigenstates of the scatterer are thus represented as sites of the tensor chain, and the MPS is completely ordered based on the corresponding energies, alternating sites describing lead and scatterer states. For Hamiltonians in the form of Eq. \eqref{kitaevham} and charging energy \eqref{Hc}, the adopted MPS sites for the scatterer are associated with its Bogoliubov quasiparticles, and we complete the system description by including an additional MPS site to account for the scatterer total charge (see Sec. \ref{ssec:charge_site} ).
Figure~\ref{fig:sketch}(b) shows a schematic representation of the MPS construction that relies on this energy basis, and the initial occupation of the lead sites is based on their chemical potential.

As a result of this construction, during the time-evolution of the system we can identify three energy intervals: for energies considerably smaller than $-V_{b}/2$ the eigenstates of the leads are approximately frozen in an occupied state, for energies larger than $V_b/2$, the eigenstates of the leads are equally frozen in the empty state; in the intermediate \textit{scattering range}, the states of the scatterer and the leads strongly interact and develop non-trivial correlations. 

\begin{figure}
\begin{center}
\includegraphics[width=\columnwidth]{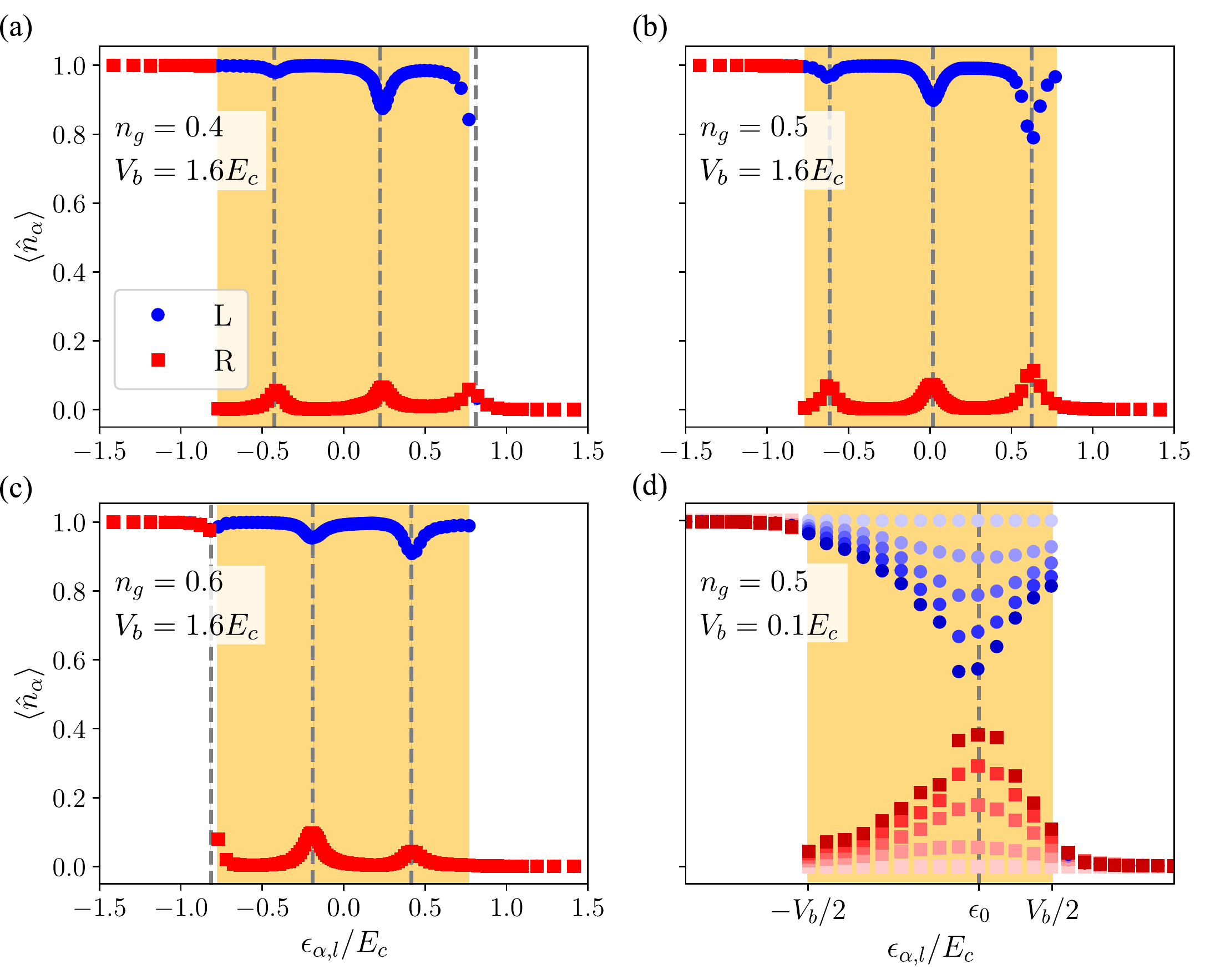}
\caption{Time evolution of SC quantum dots with two quasiparticle states of energies $\varepsilon_0=0.02E_C$ and $\varepsilon_1=0.6 E_c$. The coupling with the leads is $t_c=0.2E_c$ and $E_c=t_0$. (a-c) Occupation number $\langle \hat{n}_\alpha \rangle $ of the energy levels in the two leads as a function of their energy $\epsilon_{\alpha,l}$ after a quench, once the NEQSS is reached (${\sf t}\sim 20\hbar/t_c$). Vertical dashed lines correspond to the resonances with the transitions between different manybody states in the device. The panels refer to induced charges $n_g=0.4,0.5$ and $0.6$.
(d): time dependence of the occupation number close to a resonance for small bias $V_b=0.1E_c$. The color progression from light to dark indicates longer evolution times; the data refer to ${\sf t} =[0,20 , 40, 60, 80]$ in units of $\hbar/t_c$. Vertical dashed lines highlight the resonance at $\varepsilon_0=0.02E_c$ while the shaded areas indicate the scattering window given by the voltage bias}\label{fig:occupation_N2}
\end{center}
\end{figure}

We illustrate this behavior in Fig.~\ref{fig:occupation_N2} for a simple model of a superconducting (SC) quantum dot with two quasiparticle states and charging energy given by Eq. \eqref{Hc};
the way of dealing with the charging energy will be introduced later in Sec.~\ref{ssec:charge_site}.
Inside the scattering range (yellow shading), we can clearly observe the onset of resonances identified by the variation in time of the occupation number of the two leads at energies set by the energy differences among many-body states of the scatterer. These are identified by the vertical dashed lines in Fig.~\ref{fig:occupation_N2}(a)-(c) and such resonances are consistent with the results of the standard rate equation approaches ~\cite{Beenakker1991,Aleiner2002}.
The initial lead distributions are taken at zero temperature and the resulting width of the depletion/filling regions following the quench is set by the coupling strength $t_c$ between the leads and the SC device. Physically, this corresponds to a regime where the broadening of the device energy levels induced by the hybridization with the leads is larger than the temperature. 

The time evolution of the occupation numbers of the leads eigenstates close to a resonance is summarized in Fig.~\ref{fig:occupation_N2}(d); as time increases (from lighter to darker colors), the leads are progressively depleted/filled. The NEQSS we are interested in survives as long as there is a sufficient difference between the occupation numbers of the left and right leads.

In simulation efficiency, the advantage of adopting the energy eigenbasis is two-fold: (i) Contrary to a real state basis, most of the leads and device states are only marginally involved in the time evolution, thus the entanglement growth is limited to the scattering energy range, consistently with Ref.~\cite{Rams2020}.
(ii) This approach allows us to evaluate the conductance also for moderately large bias voltage, which is typically a difficult task for a real space basis.

Notwithstanding that the coupling between the leads and the scatterer becomes non-local in the energy basis, $\Htunn$ can be expressed in terms of matrix product operators (MPOs) of small bond dimensions~\cite{Rams2020} (in the implementation by ITensor, the MPO can be generated by AutoMPO). As we will discuss in the following, such MPO will also take into account the non-local charging energy effects.
As a consequence, the dynamics of the system can be efficiently simulated by using TDVP \cite{Haegeman_PRL2011, Haegeman2016}. In practice, we use TDVP with single-site update together with Krylov subspace expansion~\cite{Yang2020}.  The subspace expansion is performed in advance of the TDVP evolution at each time step.

We finally remark that our basis choice is opposite and complementary with respect to several other techniques, including both NRG, where typically one works in a basis such that the Hamiltonian defines an effective 1D tight-binding chain with local interactions only, and recent studies of quench dynamics in Anderson impurity models~\cite{Kohn_PRB2021,Kohn_2022}, where an alternative ordering of the MPS sites has been proposed in a position basis, in order to alternate filled and empty leads sites and simplify the description of particle-hole pairs.

\subsection{Charging energy and superconducting scatterers}\label{ssec:charge_site}

The BCS mean-field description of a superconducting system does not preserve its particle number, but only its parity. This implies that the charge $\widehat{N}$ of the scatterer cannot be simply deduced by considering the occupation of the sites of its tensor network description; therefore, to account for the charging energy interaction $\Hc$, the MPS construction must be suitably extended. To this purpose, inspired by the approach adopted in Ref. \cite{keselman2019}, we add an auxiliary bosonic site to our tensor network representation of the system (we locate it at the center of the whole MPS chain). This MPS site behaves as a counter for the number of particles in the scatterer and it is defined by a local Hilbert space spanned by the eigenstates $\ket{N}$ of the scatterer charge $\widehat{N}$. 

To keep the Hilbert space dimension of the auxiliary site finite, we introduce a truncation parameter $N_{\rm max}$, such that in our simulations we consider only a set of $2N_{\rm max} + 1$ orthogonal charge states $\ket{N}$ with $N \in \left[-N_{\rm max},N_{\rm max}\right]$ (for $n_g$ taken between -1 and 1, otherwise we can shift this range). The operator $\widehat{N}$ is diagonal in this basis and, in particular, describes the charge variation of the scatterer around the reference value $N=0$, which corresponds to the ground state of the isolated superconductor in absence of induced charge $(n_g=0)$. Given the charge truncation parameter $N_{\rm max}$, our tensor network description provides reliable results when the charging energy $E_c$ is sufficiently large, such that the fluctuations of the charge of the scatterer, determined by the coupling with the leads (and, potentially, by additional Josephson terms) are much smaller than $N_{\rm max}$. 
In our simulations, we verified indeed that for strong charging energies, thus in the case of Coulomb blockaded device, the population of states far from $N\sim n_g$ is exponentially suppressed and at the cutoff $\pm N_{\rm max}$ it remains below machine precision.

As a general rule, the introduction of such an auxiliary degree of freedom must be accompanied with a physical constraint. Here, we impose that the parity $\left(-1\right)^{\widehat{N}}$ of the auxiliary site needs to be the same as the parity of the occupation of the scatterer sites. Namely, let us define the operator:
\begin{equation} \label{parityc}
\widehat{P}= \left(-1\right)^{\widehat{N} + \sum_{i=1}^\mathcal{M} \hat{d}^\dag_i \hat{d}_i}\,.
\end{equation}
For any physical state $\ket{\psi_{\rm phys}}$, the constraint
\begin{equation} \label{constr}
\widehat{P}\ket{\psi_{\rm phys}}=\ket{\psi_{\rm phys}}
\end{equation}
must hold. Besides this physical requirement, we also observe that the total number of particles,
\begin{equation} \label{Ntot}
\widehat{N}_{\rm tot} = \widehat{N} + \sum_{\alpha=L,R}\sum_{i=1}^{\mathcal{L}} \hat{c}^\dag_{\alpha,i} \hat{c}_{\alpha,i}\,,
\end{equation}
is a conserved quantity. Our MPS and MPO constructions encode both the $\mathbb{Z}_2$ constraint \eqref{constr} and the global symmetry associated to Eq. \eqref{Ntot} through suitable quantum numbers in its virtual indices. Due to the peculiarities of the conserved operators $\widehat{P}$ and $\widehat{N}_{\rm tot}$, the local symmetry properties of the tensors associated with lead and scatterer sites are different. In particular,  the parity constraint affects only the auxiliary and the scatterer sites, whereas the particle number conservation affects only the auxiliary and the lead sites (see Appendix \ref{app:MPS} for more detail).

Based on this MPS construction, the charging energy term $\Hc$ can be included in a straightforward way in the MPO description of the Hamiltonian, since it simply corresponds to a diagonal operator acting locally on the auxiliary site.

More care is needed instead in the definition of the tunneling operators from the leads to the scatterer, which must be dressed with suitable operators acting on the auxiliary site. To this purpose, we define the operators 
\begin{equation} \label{Sigma}
\widehat\Sigma^+ = \sum_{N=-N_{\rm max}}^{N_{\rm max}-1} \ket{N+1}\bra{N} \quad {\text{and}} \quad \widehat\Sigma^- = (\widehat\Sigma^+)^{\dag} \,,
\end{equation}
 which respectively raise and lower by 1 the scatterer charge $\widehat{N}$; the tunneling Hamiltonian acquires the form:
\begin{equation}
\Htunn = -t_{c} \left[ \opddag{1}\opc{L,1}\widehat\Sigma^+ + \opddag{\mathcal{M}}\opc{R,1}\widehat\Sigma^+ +\hc \right],
\end{equation}
where the lead operators $\hat{c}$ and the scatterer operators $\hat{d}^\dag$ are taken in the real space basis. 
In the BdG basis which will be introduced below, the creation and destruction operators must be rewritten as the proper linear combination of eigenstates in the chosen representation.
Independently from the basis choice, $\Htunn$ becomes a sum of non-local three-site operators, acting simultaneously on one lead, the scatterer and the auxiliary charge site.
Despite the apparently complicated structure, the whole Hamiltonian can be efficiently represented by an MPO of bond dimension $10$. The related dynamics can thus be conveniently evaluated through a TDVP approach.

\subsection{Quasiparticle basis}\label{ssec:qpbasis}

Depending on the considered model and the corresponding $\Hsys$, we can apply different basis choices for the scatterer. When the mean-field superconducting pairing is present, for example in the SC quantum dot model analyzed in Fig. \eqref{fig:occupation_N2} or in the Kitaev chain model described by Eq. \eqref{kitaevham}, the best option for the basis is given by the quasiparticle energy eigenstates derived from the quadratic Bogoliubov - de Gennes (BdG) formulation of the corresponding Hamiltonians. Accordingly, to study such models, we developed a MPS description of the scatterer based on the occupation number of the Bogoliubov eigenmodes $\hat{\gamma}_j$ that diagonalize the BCS Hamiltonians $\Hsys$. This choice greatly speeds our simulations of the system dynamics because it avoids the strong entanglement growth caused by the formation of Cooper pairs that characterizes the real-space basis. In the single-particle energy eigenbasis, only the considerably weaker entanglement between quasiparticles contributes indeed to the entanglement entropy built during the time evolution.

In this basis, the scatterer Hamiltonian reads
\begin{equation}
    \Hsys = \sum_{j=1}^\mathcal{M} \varepsilon_j \hat{\gamma}_j^\dagger \hat{\gamma}_j - \frac{1}{2} \sum_{j=1}^\mathcal{M} (\varepsilon_j + \mu_s),
\end{equation}
where $\varepsilon_j$ and $\hat{\gamma}_j^\dagger$ are the quasiparticle energies and creation operators, corresponding to the \emph{positive} eigenenergies of the BdG Hamiltonian. The first and the second term correspond to the excited states and the ground state energies respectively.
In this basis, the real-space operators $\hat{d}_i$ acquire the form $\hat{d_{i}} = \sum_{j=1}^\mathcal{M} u_{ij} \hat{\gamma_j} + v_{ij}^* \hat{\gamma_j}^\dag$.
We stress that although the complete BdG Hamiltonian has $2\mathcal{M}$ eigenstates, half of them with positive energies and half of them with negative energies, the $\mathcal{M}$ positive energy states are enough to form a complete set due to particle-hole symmetry.
Therefore the number of basis states in the new basis remains unchanged.

As discussed in the previous section, the first step in the simulation of the time evolution of the system for the density quenches requires to determine the ground state of the scatterer.
Since $\Hsys$ and $\Hc$ are completely decoupled in our representation, this ground state is the tensor product of two parts referring to the Bogoliubov quasiparticles and the auxiliary charge site.
 Such tensor product can display either even or odd fermionic parity and it respectively assumes the form $|\psi^+\rangle|N=0\rangle$ or $|\psi^-\rangle|N=1\rangle$, for $n_g \in \left[0,1\right]$. These are indeed the combinations that fulfill the physical constraint in Eq. \eqref{constr}.

In the quasiparticle basis, $|\psi^+\rangle$ and $|\psi^-\rangle$ are simply the vacuum state and the state $|1,0,0, \ldots\rangle$ respectively; for general bases and $\Hsys$, $|\psi^\pm\rangle$ can be computed by DMRG. The ground state is then determined as the state of lower total energy including both the contributions of $\Hsys$ and $\Hc$, and it is adopted to initialize the time evolution.

\begin{figure*}
\begin{center}
\includegraphics[width=0.55\columnwidth]{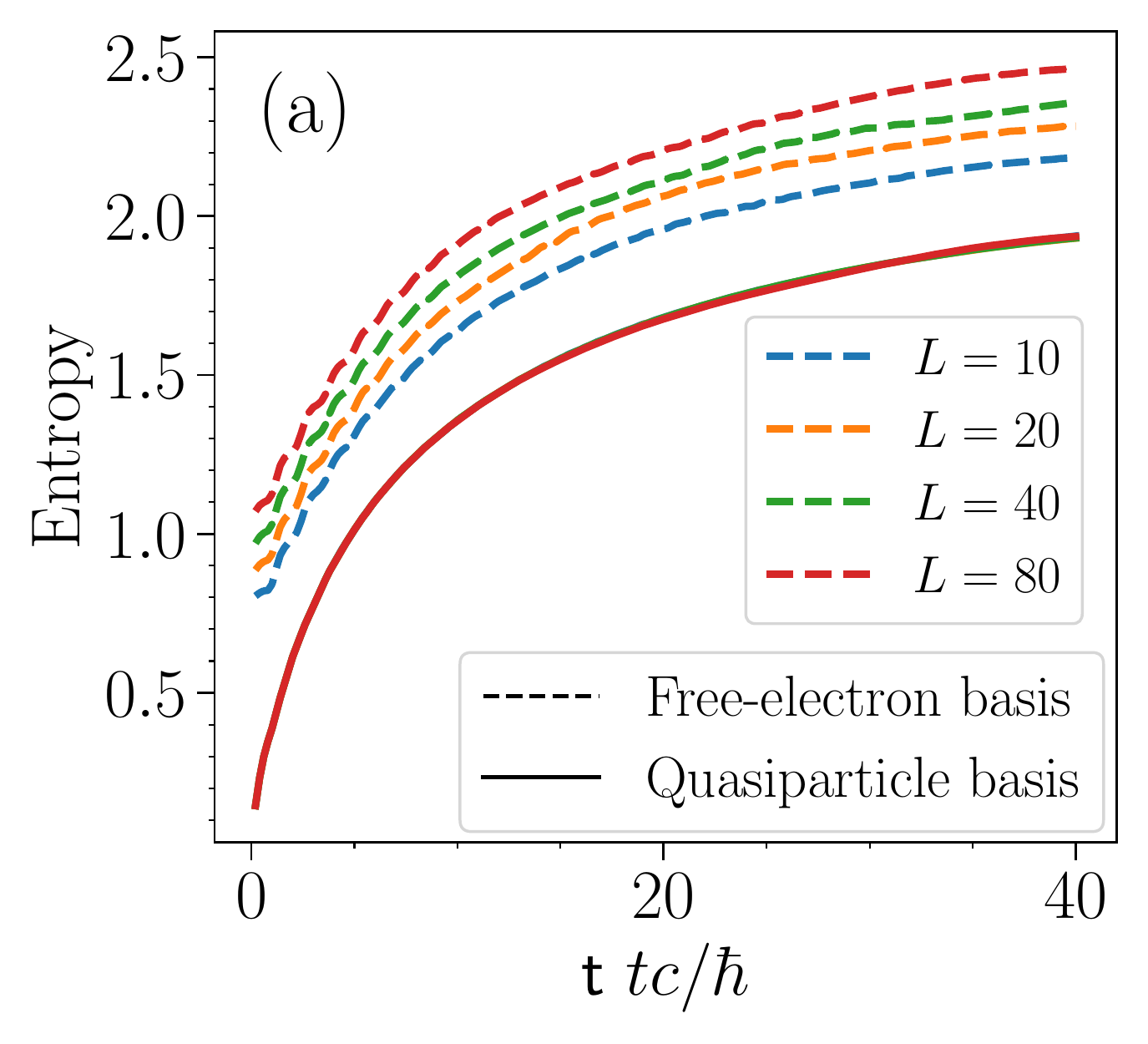}
\includegraphics[width=0.55\columnwidth]{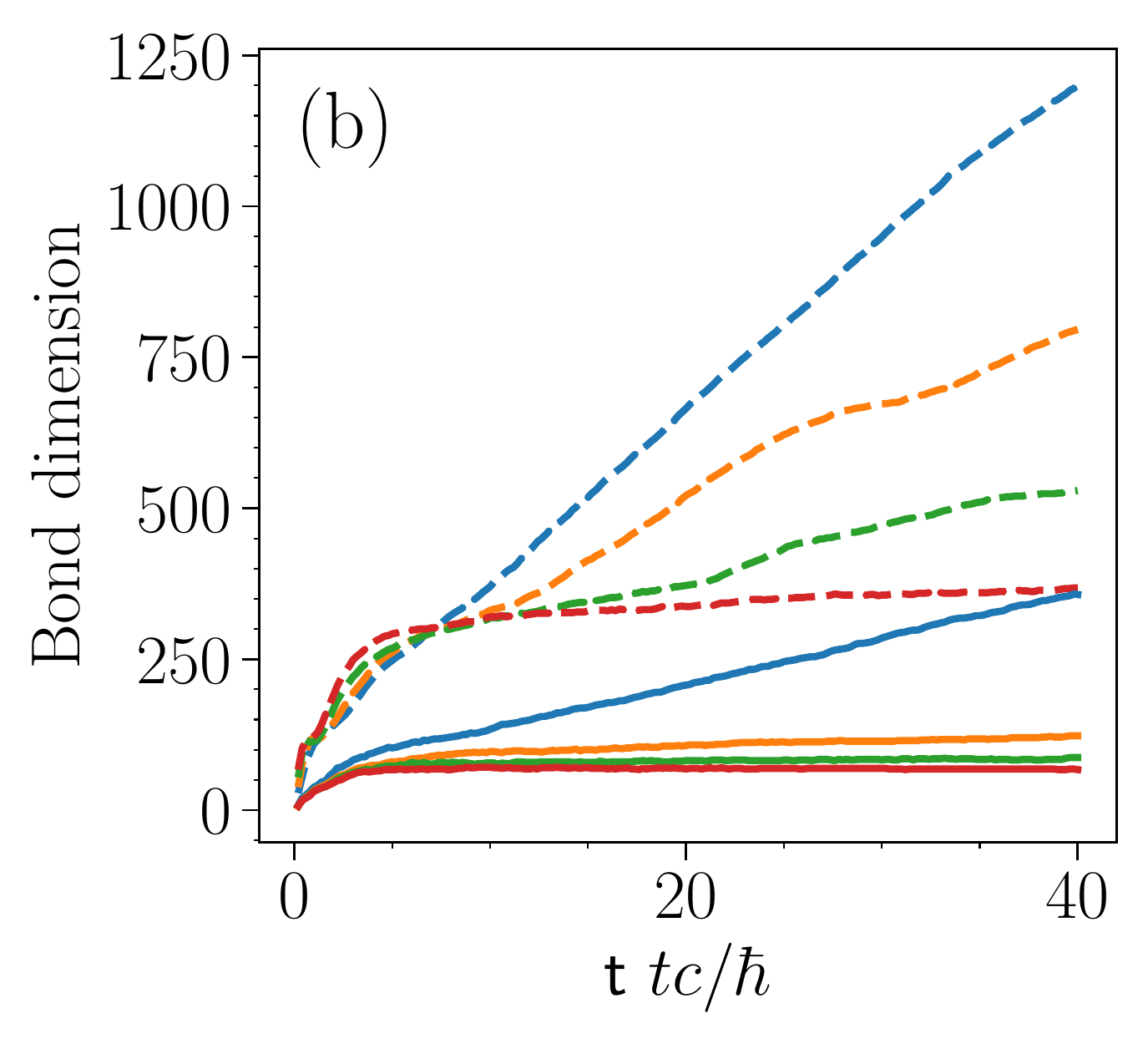}
\includegraphics[width=0.4\columnwidth]{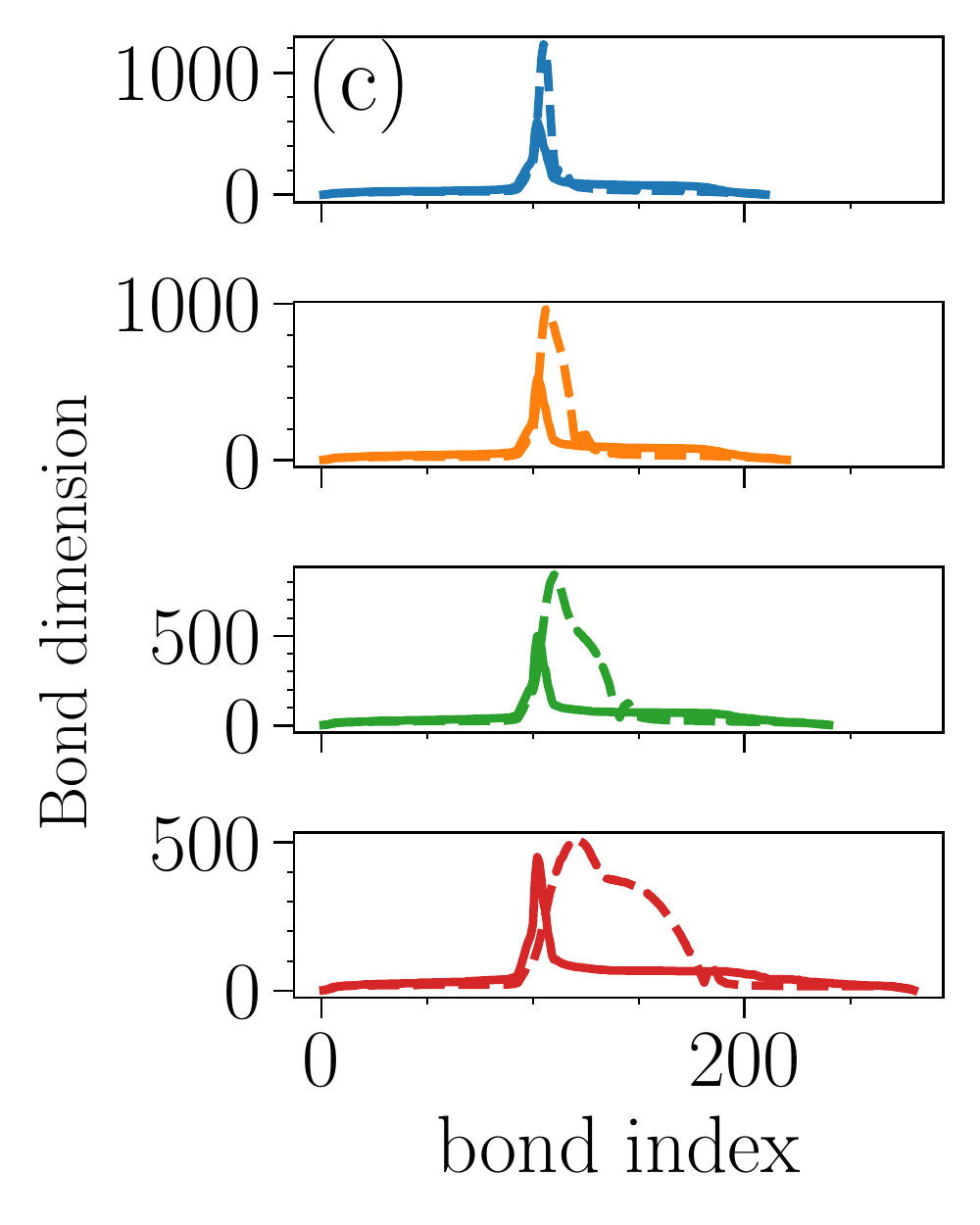}
\includegraphics[width=0.5\columnwidth]{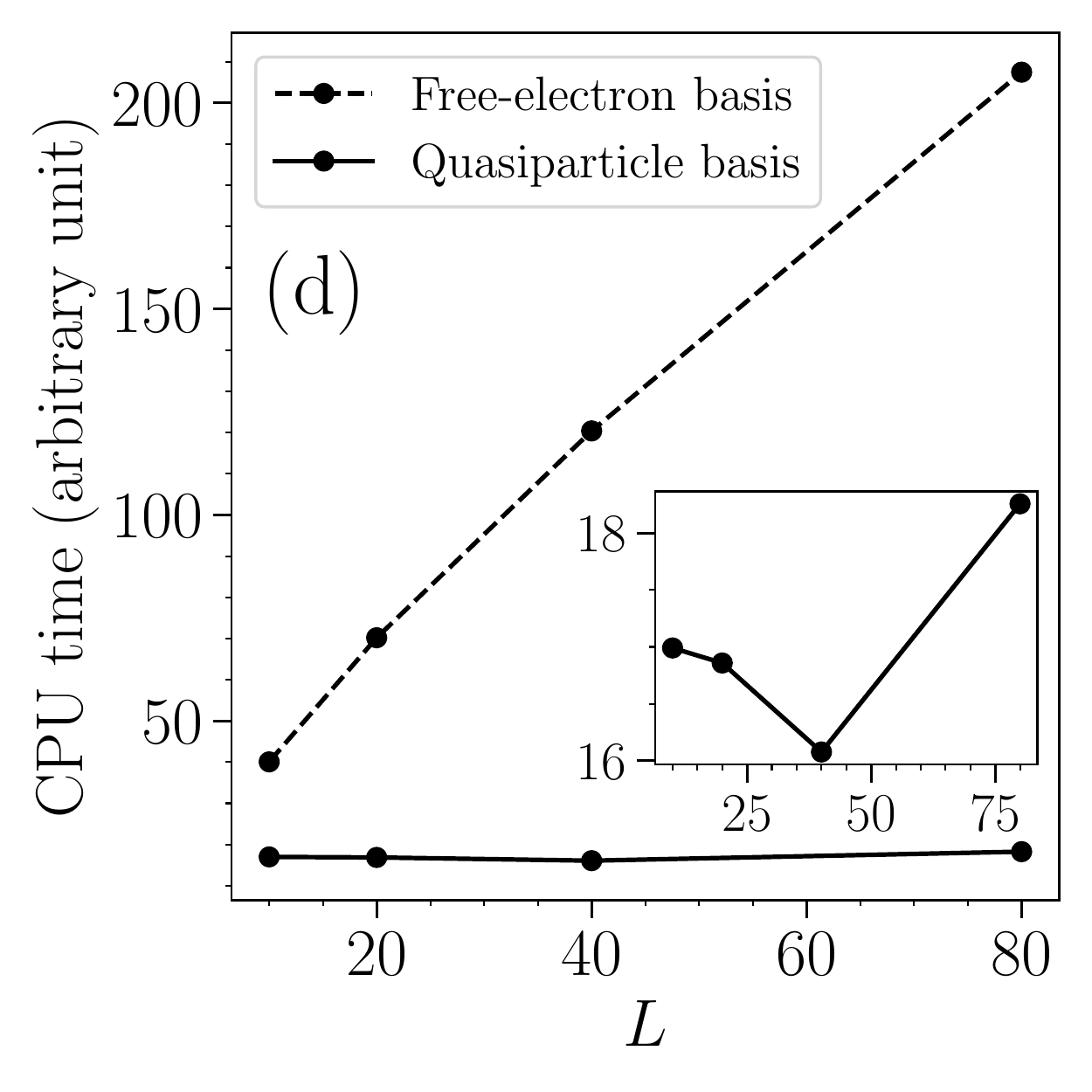}
\caption{
Comparisons for quench dynamics between a free electron basis (dashed lines) and the Bogoliubov quasiparticle basis (solid lines). The scatterer is a Kitaev chain with parameters $E_c=t_0$, $t_s=0.4E_c$, $\mu_s =-0.1E_c$, $\Delta=0.5E_c$, $t_c=0.2E_c$, $n_g=0.5$, and $\Vbias=0.1 E_c$, the same with Fig.~\ref{fig:current_example}.
(a) Maximal entanglement entropy and (b) largest bond dimension reached in the time evolution. (c) Bond dimension on each bond in the MPS. (d) CPU time for the whole simulations for different scatterer lengths. The inset is a zoom in of the solid curve.
}
\label{fig:basis_compare}
\end{center}
\end{figure*}

To illustrate the efficiency of the quasiparticle basis, we consider a Kitaev chain in the topological phase and compare the convergences to a NEQSS under small voltage bias obtained by using two different basis for the MPS construction:
the first corresponds to the free electron eigenbasis of the kinetic term $t_s$ only [see Eq. \eqref{kitaevham}]; in this case, the pairing term is included in the MPO; the second corresponds instead to the Bogoliubov quasiparticle basis, which already takes into account the superconducting pairing.
As shown in Fig.~\ref{fig:basis_compare}(a), for the free electron basis, the entanglement entropies become progressively larger for longer systems (dashed lines). Instead, for the Bogoliubov quasiparticle basis, the entanglement entropies are basically independent on the system length, and are smaller than the entropies in the free electron basis.
Surprisingly, the largest bond dimensions required by the algorithm during the time evolution become smaller for longer chains in both bases, as shown in Fig.~\ref{fig:basis_compare}(b). This can be understood as an effect of the finite overlap acquired by the Majorana edge-modes for short chains, which enhances the entanglement.
For all lengths, the largest bond dimensions reached in the Bogoliubov basis are smaller than the bond dimensions in the free electron basis.
Another advantage of using the quasiparticle basis can be seen in the bond dimension distribution on each bond in the MPS.
As shown in Fig.~\ref{fig:basis_compare}(c), the region of large bond dimension is broader in the free electron basis than in the quasiparticle basis.
This is because under small voltage bias the transport is mainly contributed by the Majorana bound state (MBS), which corresponds to a single site in the quasiparticle basis, whereas in the free-electron basis, the MBS is an entangled state spread through the whole scatterer region.
Fig.~\ref{fig:basis_compare}(d) shows the comparison of CPU times. The Bogoliubov quasiparticle basis is in general more efficient than the free electron basis, and the advantage becomes stronger when the system length is larger.

The construction we presented can be extended, in full generality, by considering any arbitrary quasiparticle basis for the device, and considering a suitable tunneling matrix between the leads and the scatterer. In particular, for the modeling of the transport across specific nanostructures, one may want to consider only a selected number of low-energy single-particle eigenstates by including only the most relevant states involved in the transport process, rather than starting from a tight-binding microscopic model such as Eq.~\eqref{kitaevham}.
This kind of approximation has been successfully applied, for instance, to study Coulomb blockaded transport within the framework of rate equations~\cite{Higginbotham_NatPhys2015,Vaitiekenas_PRB2022}; 
our MPS representation can easily be adapted to this approach and, together, they also provide a systematic way of improving the resulting conductance estimates by increasing the number of low energy states included in the scatterer description.

\subsection{Josephson energy and inclusion of a superconducting lead}\label{ssec:Josephson}

So far, we discussed systems in which the scatterer exchanges particles with the leads only. The introduction of the auxiliary site, however, allows us to to extend further our model in order to effectively account for an additional SC lead which exchanges Cooper pairs with the SC scatterer. This is possible by replacing the Hamiltonian $\Hc$ with a more general form:
\begin{equation} \label{SCbox}
\widehat{H}_{\rm SC \, box} = E_c\left(\widehat{N} - n_g\right)^2 - E_J \left[ (\widehat{\Sigma}^+ )^2  + (\widehat{\Sigma}^- )^2\right]\,.
\end{equation}
This constitutes the Hamiltonian of a Cooper pair box \cite{Bouchiat_1998}, and the energy scale $E_J$ represents its Josephson energy. The second term in Eq. \eqref{SCbox} varies the number of electrons in the scatterer by $\pm2$ [see Eq. \eqref{Sigma}] and indeed represents a process in which Cooper pairs tunnel in and out of the system. Given the limitation provided by the truncation $N_{\rm max}$, the simulation of the system is reliable only when $E_J$ is sufficiently smaller than $E_c$, such that, in practice, the states $\ket{\pm N_{\rm max}}$ display a negligible population. Therefore, the extension of the original model provided by Eq. \eqref{SCbox} is suitable to describe scatterers which are typically in a Coulomb-dominated regime, as in the case of SC charge qubits. The opposite transmon limit $E_J > E_c$, instead, cannot be satisfactorily explored.

When replacing $\Hc$ with $\widehat{H}_{\rm SC \, box}$, the conservation of total particle numbers (in the leads and the scatterer) is reduced to the conservation of parity, which may assume both values depending on the initial lead states.

By including the Josephson energy, the model becomes effectively a three-terminal device, able to simulate the interacting Coulomb blockaded regime of systems analogous to the ones studied in Refs. \cite{Danon2020,Menard2020}. We also observe that, by switching off the tunnel coupling with one lead, it is possible to study two-terminal normal-superconducting junctions mediated by a blockaded scatterer described by $\Hsys$.

\section{Physical examples}\label{sec:benchmark}
In this section we benchmark our method on two test systems: a p-wave SC quantum dot with two quasiparticle states and a Kitaev chain.
We will use the Kitaev chain to illustrate the main features of our method as well as its limitations, whereas the small Hilbert space dimension of the quantum dot allows for faster simulations and an easier characterization of the Coulomb blockaded structure.

In general, we simulate the nonequilibrium dynamics following a density quench: For any choice of $n_g$ and $\Vbias$, the system is initially prepared in the corresponding ground state with no couplings between the leads and the device ($t_c=0$).
At ${\sf t}=0$, we turn the tunnel coupling $t_c$ on and quench the voltage bias to zero; the transport properties of the system are then estimated based on the NEQSS reached after the initial transient time. Throughout all the simulations, the main observable we measure is the current flowing from one lead to the other as a function of the voltage bias $\Vbias$ and the induced charge $n_g$.
The current is measured on both edges of the central device, in the first links {\em entirely} in the leads [$l=1$ in Eq. \eqref{currentop}], to facilitate comparison between data associated to different couplings between the leads and the scatterer or hopping decay length in the leads.
Since we always consider superconducting models and symmetric voltage drops, we assume that the current $I(\Vbias, n_g)$ is an odd function of the the voltage bias, hence we perform all simulations for $\Vbias >0$, corresponding to a particle current flowing from left to right.
After the current has converged to the stationary value,  we divide each data set in several batches from which we compute its average value and standard deviation.
From the current we compute the differential conductance $G=\frac{{\rm d}I}{{\rm d} \Vbias}$ using a fourth-order discrete derivative method.
Other meaningful observables we can extract are the charge on the SC island and the entanglement entropy on each link, which is naturally obtained during the singular value decomposition (SVD) performed at each step of the time evolution.

\begin{figure}
  \begin{center}
    \includegraphics[width=\columnwidth]{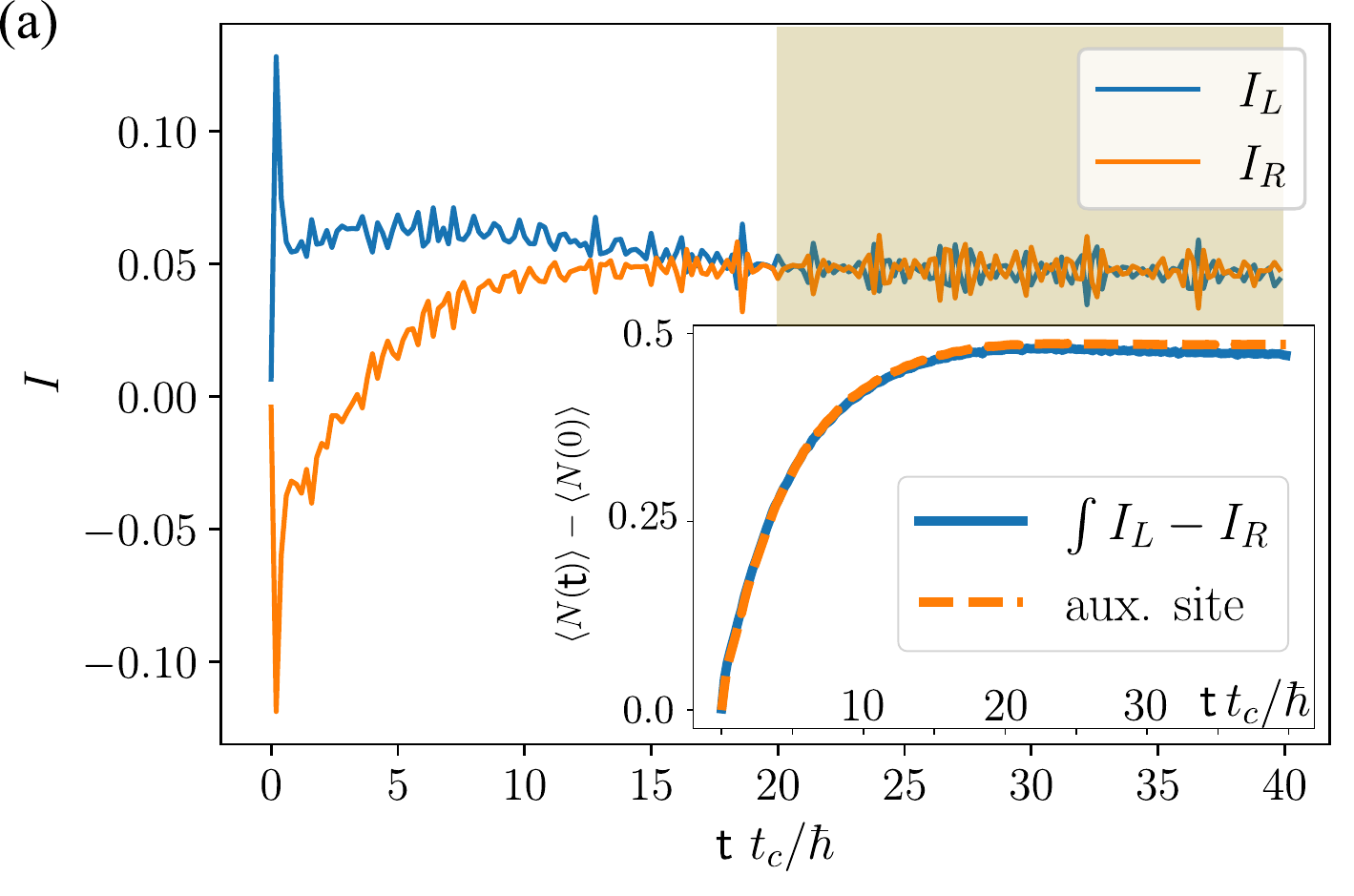}    
    \includegraphics[width=\columnwidth]{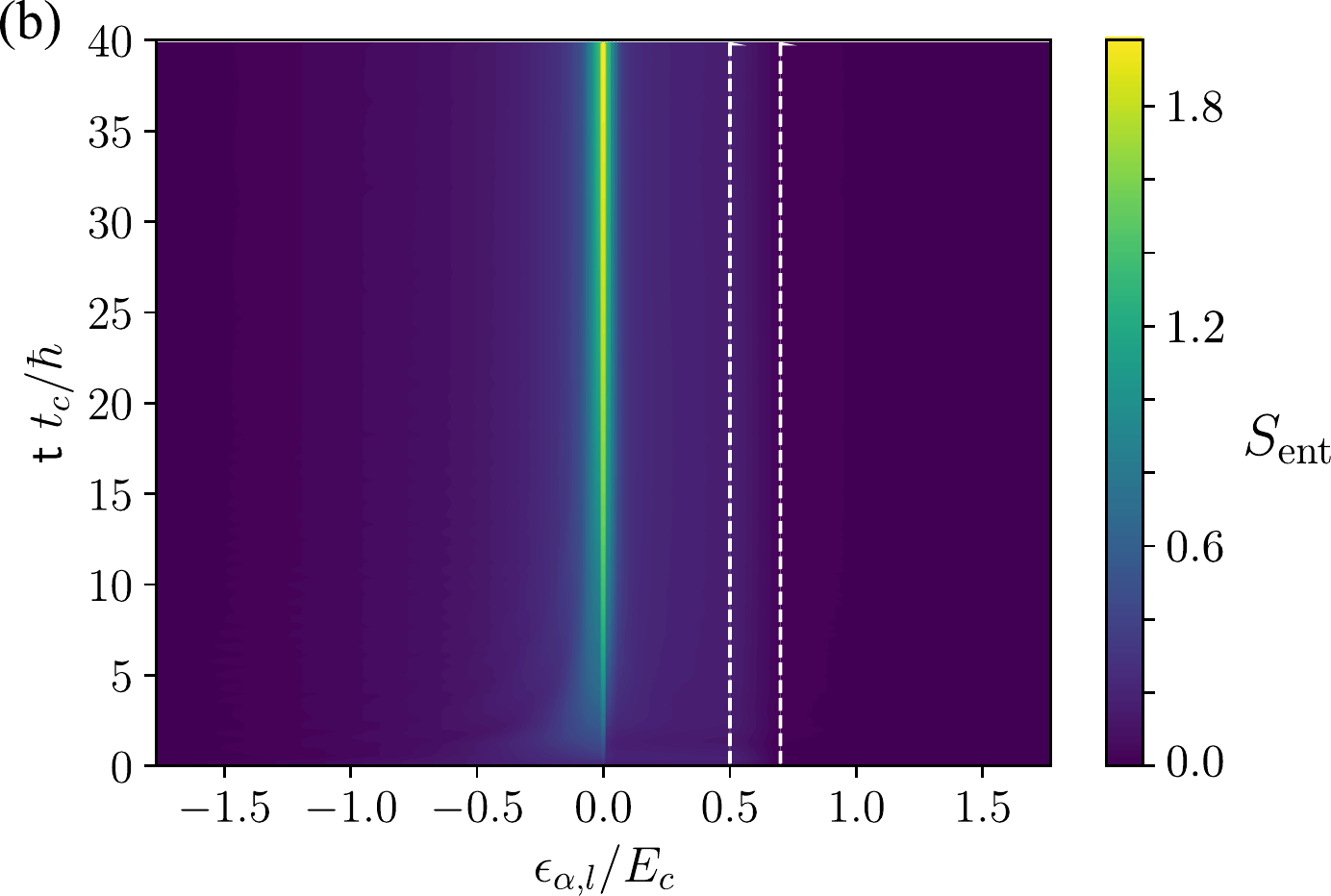}
    \caption{(a) Evolution in time of the left and right currents ($I_L$ and $I_R$ respectively) after the density quench. The shaded area corresponds to the interval where we average the current to extract its steady-state value. 
    The inset shows the change in time of the charge of the SC island, both as the average particle number in the auxiliary site (dashed line) and as the integral of the current difference between the left and right contact (solid line). 
  (b) Entanglement entropy at each bond of the MPS chain. The two vertical dashed lines indicate the region where the excited quasiparticle states of the Kitaev chain are located, while the auxiliary charge site and the Majorana modes lies at zero energy, where most of the entanglement is concentrated.
    The parameters in the Kitaev chains are $E_c=t_0$, $t_s=0.4E_c$, $\mu_s =-0.1E_c$, $\Delta=0.5E_c$, and $t_c=0.2E_c$.
    The current is computed in proximity of the zero-bias peak at the charge degeneracy point $n_g=0.5$, $\Vbias=0.1 E_c$.}\label{fig:current_example}
  \end{center}
\end{figure}

To give an example of the behavior of these quantities in the quench protocol, we report in Fig.~\ref{fig:current_example} the results for a topological Kitaev chain of $\mathcal{M}=40$ sites at the charge degeneracy point $n_g=0.5$ and a small voltage bias $\Vbias=0.1E_c$. Panel (a) reports the postquench time dependence of the current at the left and right edges of the scatterer;
its behavior is very similar to that shown in the noninteracting case, illustrated in Fig.~\ref{fig:quench_exact}, suggesting that the MPS simulation is capturing correctly the transport phenomenon. 
The inset shows the corresponding behavior of the charge accumulated on the device, comparing the difference between the ingoing and outgoing current and the variation of the occupation number of the auxiliary charge site. 
Their agreement is a good sanity check to verify that the simulation is physically meaningful and the stationary value $\langle \hat{N}({\sf t}) \rangle -\langle \hat{N}(0) \rangle \sim 0.5$ is correctly reached:
The system is initially prepared in the BCS ground state with no quasiparticle excitation ($\langle \hat{N}(0) \rangle=0$) which is degenerate with the state with the MBS occupied (since we are deep in the topological phase); 
after the quench, the system approaches a NEQSS characterized by an equal superposition of the two degenerate many-body states, hence the increase of the charge.
Panel (b) shows the spread of the entanglement in the energy-ordered MPS chain: thanks to the basis choice, the entanglement is confined to low energy states. In particular, states of the Kitaev chain above the energy gap ---located in the region between the two dashed vertical lines--- remain almost uncorrelated with the rest of the system, signaling that transport is mediated mainly by the Majorana modes.
This situation represents the ideal case of application of our method: when only a few scatterer states are involved in the transport process, while most remain untouched by the dynamics, the convergence of the simulation towards a NEQSS is robust and it is not limited by the system size.

\subsection{SC quantum dot}\label{ssec:qdot}
\begin{figure*}[ht]
\begin{center}
\includegraphics[width=18cm]{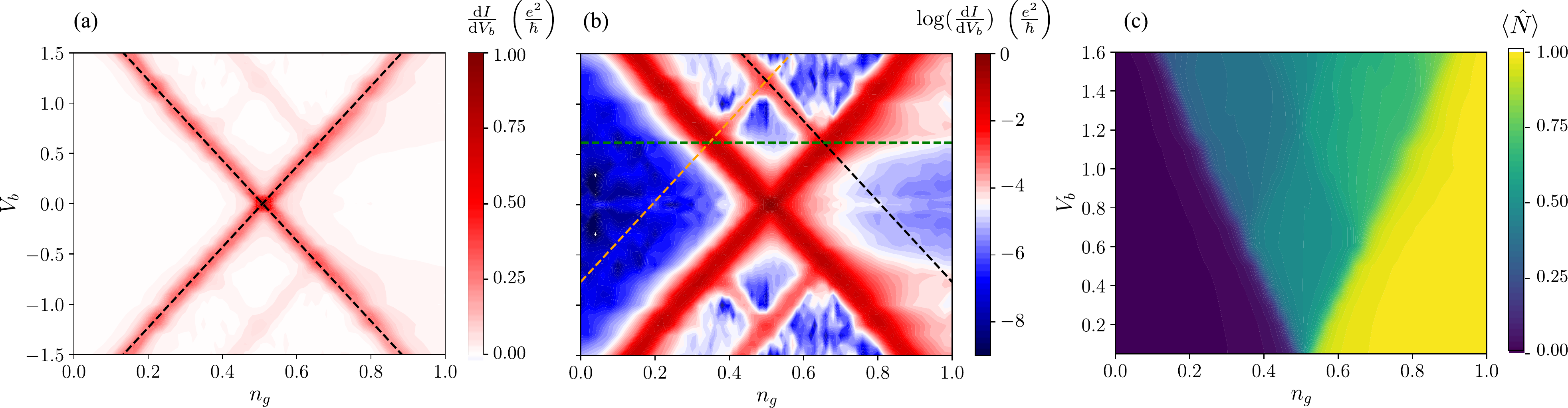}
\caption{(a) Differential conductance close to a charge degeneracy point in the $n_g$-$V_b$ plane, for a two level SC systems with $\varepsilon_0=0.02 E_c$, $\varepsilon_1=0.6 E_c $, and $t_c=0.2 E_c$. Transport is dominated by the sequential tunneling resonances (black dashed lines), corresponding to $V_b/2 = \pm [ E_c(1-2 n_g)+\varepsilon_0 ]$.
(b) Differential conductance in log-scale, illustrating the cotunneling step in the odd Coulomb valley that arises at $V_b \sim \pm (\varepsilon_1-\varepsilon_0)$ (green dashed line).
Also the transitions involving excited states are more visible: $V_b/2= [ E_c(1-2 n_g)+\varepsilon_1 ]$ (black dashed line) and $V_b/2= [ E_c(2 n_g-1)+\varepsilon_1 ]$ (orange dashed line).
(c) Average charge in the NEQSS of the SC quantum dot, for positive bias. Notice how the sequential tunneling resonances are reflected in the structure of $\langle \widehat{N} \rangle $.  }\label{fig:SCdot}
\end{center}
\end{figure*}

A clear example of the physical results that can be explored by our approach is provided by a floating p-wave superconducting two-level system, one of the simplest SC model that displays Coulomb blockaded transport. 
Its small size allows for fast and accurate simulations over a wide bias range, making it easy to characterize the Coulomb blockaded differential conductance in the whole $n_g-\Vbias$ plane.
Indeed, the limited growth of the entanglement in this model yields that there is no need of fine tuning the simulation parameters (as the duration of the TDVP time steps or the decay length $\xi$) to ensure the convergence to a NEQSS.

This system is represented by the Hamiltonian:
\begin{equation}\label{eq:twolevel}
\Hsys = \varepsilon_0 \opgammadag{0}\opgamma{0} + \varepsilon_1\opgammadag{1}\opgamma{1} \ , 
\end{equation}
where $\opgamma{1,2}$ are the destruction operators of the quasiparticle levels \footnote{On a practical level, the SC quantum dot is described as a very short Kitaev chain \eqref{kitaevham} with $\mathcal{M}=2$ sites only.}.
Although simple, this toy model can be used as an approximation for a hybrid semiconductor/superconductor nanowire with strong spin-orbit coupling to allow for the emergence of an effective p-wave SC pairing (see, for instance, Refs. \cite{Higginbotham_NatPhys2015,Vaitiekenas_PRB2022}): in this case, the lowest energy level $\varepsilon_0$ represents a non-degenerate subgap state (Andreev or Majorana bound state), whereas the eigenstate at $\varepsilon_1\sim \Delta$ constitutes an effective representation of all the quasiparticle states above the SC gap. 
To reconstruct the Coulomb blockade diamonds, we perform the quench simulations of a grid of points in the $n_g$-$\Vbias$ plane, with a denser sampling close to the zero-bias peak, and compute the time-dependent current averaged over the left and right contacts. 
Then, we follow the procedure described at the beginning of Sec.~\ref{sec:benchmark} to extract the differential conductance.

Our results show that the MPS+TDVP simulation allows to capture all the expected perturbative transport features of these SC blockaded systems, including both sequential tunneling resonances and inelastic cotunelling effects at finite bias. In particular, Fig,~\ref{fig:SCdot}(a) illustrates the differential conductance in the $n_g$-$\Vbias$ plane for a quantum dot with $\varepsilon_0 = 0.02 E_c$ and $\varepsilon_1 =0.6 E_c$. The charging energy is the dominant energy scale, alongside the ``bare'' hopping amplitude in the leads $t_0=E_c$, while the tunnel coupling between the leads and the system is $t_c=0.2 E_c$. 

The first feature emerging in the differential conductance is the appearance of resonances caused by the incoherent sequential tunneling mediated by the low energy state $\varepsilon_0$.
These bright conductance resonances are clearly visible [indicated by black dashed lines in Fig.~\ref{fig:SCdot}(a)] and appear when the voltage bias matches the energy difference between the ground states in the even and odd sectors $\Vbias/2 = \pm E_c(1-2n_g)+\varepsilon_0$. Sequential tunneling is indeed the main transport mechanism emerging in the perturbative rate equation approaches and the related conductance peaks are commonly observed in superconducting blockaded devices; see, for instance, the experimental data referring to SC islands in nanowires in Refs. \cite{Higginbotham_NatPhys2015, Albrecht2016, Vaitiekenas_PRB2022, Vekris_DNW2021} and the theoretical analysis in \cite{vanHeck_PRB2016,Lai_PRB2021}.

The sequential tunneling mediated by the state $\varepsilon_0$ is not the only perturbative feature that characterizes the intermediate $t_c$ regime we are exploring. In Fig.~\ref{fig:SCdot}(b) we report the same data in logarithmic scale, where a richer structure emerges more clearly. One can indeed observe the fainter resonances corresponding to sequential tunneling processes involving the high-energy quasiparticle level $\varepsilon_1$ (orange and black dashed lines). 

Interestingly, besides these sequential tunneling features, which correspond to first-order phenomena in the tunnel coupling $t_c$, we can clearly spot cotunelling effects, related instead to second-order phenomena in $t_c$. The main cotunelling feature is the appearance of an even-odd effect that distinguishes Coulomb diamonds with  different parities of the particle number $\widehat{N}$ ~\cite{vanHeck_PRB2016,Vaitiekenas_PRB2022}. 
In particular, the finite-bias conductance in the Coulomb valleys with odd particle number [$n_g >0.5$ in Fig. \ref{fig:SCdot}(b)] displays an inelastic cotunneling step visible for $\Vbias \sim \varepsilon_1 - \varepsilon_0$ (green dashed line).
The lack of appreciable cotunneling in the even valley is due to the destructive interference between the possible cotunneling paths when the ground state has even parity, while the interference becomes constructive in the odd valley.
Indeed, with our method we simulate the coherent evolution of a closed quantum systems, where interference effects can play a dominant role. In their absence, cotunneling steps would appear also in the even valley; in particular, when the lowest quasiparticle state has an energy $\varepsilon_0 \ll E_c,\  \Delta_{SC}$, the conductance predicted by rate equations in the even and the odd valley would be almost identical.

Finally, in Fig.~\ref{fig:SCdot}(c), we report the average occupation number $\langle \widehat{N} \rangle$ of the auxiliary site that describes the total charge of the device.
In the simulation we set the charge truncation at $N_{\rm max}=5$.
This truncation has very little effect on the simulations since in the steady state the average charge acquires values between 0 and 1, for $n_g \in [0,1)$, while the occupation of the states at the cutoff $\widehat{N}=N_{\rm max}$ lies below numerical precision.
From the figure, it is clear that, when transport is suppressed, the SC island has a well defined integer charge determined by $n_g$ [blue and yellow areas in Fig.~\ref{fig:SCdot}(c)].
When transport is present, instead, the device is in a mixed state resulting in an average charge $\langle \widehat{N} \rangle \sim 0.5$, with a series of plateaus delimited by the sequential tunneling resonances. 

Despite the fact that both sequential tunneling and inelastic cotunneling features are evident in our simulations, we cannot expect a quantitative agreement on the amplitude of the differential conductance peaks between our non-perturbative simulations and the perturbative rate equation techniques.
There are indeed two aspects to be emphasized. First, our MPS calculations simulate the unitary evolution of a closed system at zero temperature, whereas rate equations only describe the non-coherent evolution of the populations of the scatterer many-body states. Therefore the MPS+TDVP method captures interference effects between transport channels in higher-order processes such as cotunneling, as mentioned before, while rate equations do not. Second, rate equations are rigorously justified only when the temperature is larger than the tunnel coupling $t_c$ and mainly describe temperature-broadened conductance peaks (see App.~\ref{app:rates}). In this respect the tunneling strength we used for most of the simulations, $t_c = 0.2 E_c$, is definitely beyond their range of validity.

Finally, we remark that the MPS approach is able to capture also non-perturbative effects. In this respect, we observe that a faint zero-bias peak is visible in Fig. \ref{fig:SCdot}(b) for $0.5 <n_g <0.9$, thus in the odd diamond. Its appearance suggests the onset of a weak non-perturbative Kondo-like effect which cannot be captured by rate equations.

Therefore, our technique provides a method which is complementary to the standard rate equation approach: the MPS+TDVP simulations allow for the investigation of transport in the low-temperature and strong coupling regime, which is typically hard to tackle with traditional techniques. For small tunneling rates and higher temperatures, instead,  transport is dominated by perturbative phenomena which can be efficiently captured by rate equations.

\subsection{Coulomb blockaded Kitaev chain}\label{ssec:kitaev}

We now analyze the quantum transport across a Kitaev chain in the topological phase and characterize the zero-bias peak of the differential conductance and its dependence on the voltage bias at the charge degeneracy point. 
By simulating the dynamics of a chain with $\mathcal{M}=40$ sites, we show that our method correctly captures the low-bias transport mediated by the Majorana modes. 
The analysis of the height and width of the conductance peak as a function of the tunnel coupling $t_c$ shows that we can investigate the dynamics of large systems in a strong coupling regime, well beyond the validity of both perturbative rate equations and single resonant level~\cite{vanHeck_PRB2016} approaches.
To characterize the current dependence of the voltage bias at the charge degeneracy point $n_g=0.5$, we focus, instead, on a shorter chain;
even though we can study much longer {\em gapped} systems at {\em small biases}, we chose $\mathcal{M}=8$ because the simulations with a considerably larger number of quasiparticle states and large biases are subject to a fast entanglement growth and their correlations rapidly saturate the maximum bond dimension we set.

The most characteristic feature of Coulomb blockaded transport is the zero-bias peak that appears at the charge degeneracy point. 
Its presence is easily understood by a first order rate equation approach (see App.~\ref{app:rates}): when the energy of the BCS ground state with even fermionic parity -- i.e. no quasiparticle (QP) -- matches the energy of the ground state with odd parity -- thus with the lowest lying QP state occupied -- electrons can tunnel into and out the SC device without paying energy.
In the presence of Majorana zero-energy modes, however, this process results in a coherent teleportation of electrons between the two leads mediated by the MBS \cite{Sodano2007,Fu2010}. 
Therefore, in the limit of large energy separation between the zero-energy MBS and the other QP states, the zero-bias conductance peak can be estimated based on a non-perturbative Breit-Wigner (BW) formula for resonant tunneling mediated by the Majorana modes~\cite{vanHeck_PRB2016}.
Indeed, if we consider a single resonant fermionic level, the zero-bias differential conductance close to the charge degeneracy point $n_g=0.5$ is approximated by 
\begin{equation}\label{eq:BW}
G_{BW} = \frac{e^2}{h} \frac{\Gamma_L \Gamma_R}{4E_c^2(n_g-0.5)^2+(\Gamma_L+\Gamma_R)^2/4} \ ,
\end{equation}
where $\Gamma_{\alpha} = t^2_{c,\alpha}|u_\alpha|^2/t_0$ is the effective tunnel rate that takes into account the local density of states $\nu = (2 \pi t_0)^{-1}$ of the leads with open boundary conditions at the Fermi energy  and the projection of the particle-like component of the resonant level on the device edges, $u_\alpha$.  $G_{BW}$ neglects the transport effect of QP states above the SC gap and results in the quantization of the differential conductance peaks for symmetric rates $\Gamma_R=\Gamma_L$.

In the following, we compare our numerical results with this BW theoretical prediction. In analogy with Eq. \eqref{eq:BW}, also the MPS simulations are not perturbative in the tunnel coupling $t_c$; at the same time, they provide a more complete description than the BW formula because they take into account the presence of multiple energy levels above the SC gap.
\begin{figure}
  \begin{center}
    \includegraphics[width=\columnwidth]{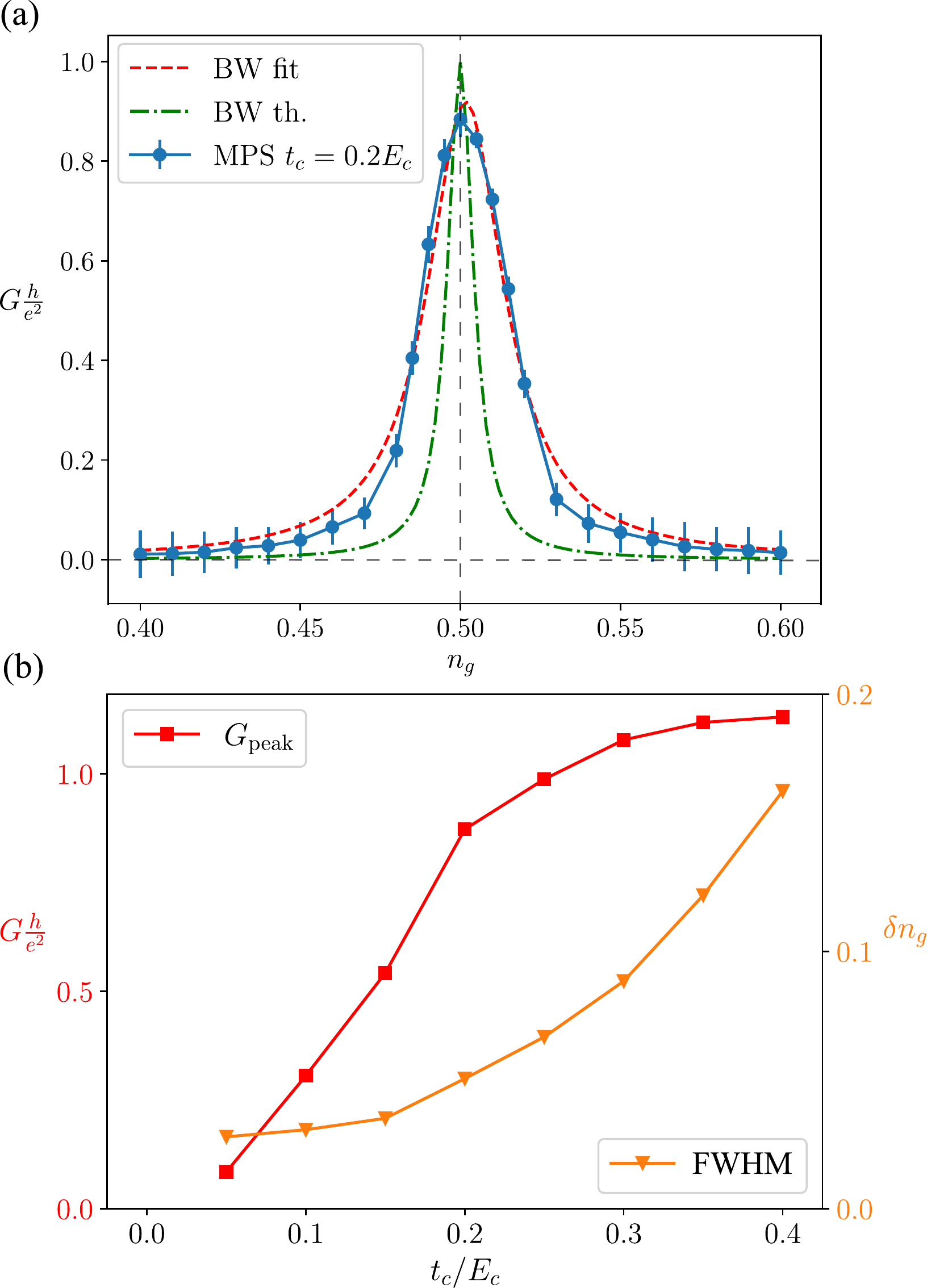} 
    \caption{(a): Zero-bias Coulomb peak of the differential conductance $G=\frac{{\rm d}I}{{\rm d}\Vbias}$ close to the resonance at $n_g=0.5$; we compare the results from the MPS + TDVP simulation (blue circles) for symmetric couplings $t_{c,L}=t_{c,R}=0.2 E_c$, the BW formula Eq.~\ref{eq:BW} (dot-dashed curve), which predicts the sharp quantization of $G$, and a fit using again Eq.~\eqref{eq:BW} (dashed line), where the height and width of the peak are used as fit parameters.
The other physical parameters are $\Delta=t_{\rm s}=0.3E_c$, $\mu_{\rm s}=-0.1E_c$, $E_c=t_0$ and $\mathcal{M}=40$. 
(b): height (red squares, left vertical axis) and the full width at half maximum (FWHM, orange triangles, right vertical axis) of the conductance peaks as a function of the tunnel coupling $t_c$. 
}\label{fig:peak}
  \end{center}
\end{figure}

In Fig.~\ref{fig:peak}(a), we compare the zero-bias peak in the differential conductance $G=\frac{\ud I}{\ud \Vbias}$ obtained from the MPS simulation for symmetric couplings with the two leads (blue curve), Eq.~\eqref{eq:BW} (green dash dotted curve), and a fit of the numerical data with a generic BW function in which the height, broadening and center are considered as free parameters (red dashed curve).
Despite a correct overall peak shape, our simulations do not recover the quantized conductance predicted by Eq.~\eqref{eq:BW} when the left and right couplings are equal; furthermore the peak's width is about twice as large as the bare BW prediction of Eq.~\eqref{eq:BW}.
Indeed, the data in Fig.~\ref{fig:peak} refer to a strong coupling scenario, where $t_c=0.2E_c \sim 0.4 \Delta_{SC}$, where $\Delta_{SC}$ is the energy separation between Majorana and excited states. In this case, the hybridization strength is not negligible and the quench dynamics is sensitive to the presence of the higher energy levels.
Moreover, the quantization of the conductance predicted by Eq.~\eqref{eq:BW} stems from an approximation based on leads with infinite bandwidth and a constant density of states, two limits which are clearly not attained in our approach.
Notice also that the peak's positions of our data and of the corresponding fit are slightly shifted to the right with respect to the bare BW result in Eq. \eqref{eq:BW}. 
This can be interpreted as an effect of the strong hybridization between the leads and the Kitaev chain, which might add a small energy shift to the energy of the Majorana modes.

When changing $t_c/E_c$, both the height ($G_{\rm peak}$) and the full width at half maximum (FWHM) of the conductance peak change, as shown in Fig.~\ref{fig:peak}(b). 
Both decrease when the coupling becomes smaller. In particular, we observe a sublinear growth of the maximum conductance with $t_c$, for $t_c\gtrsim 0.2 E_C$, which is in contrast with both the first order perturbation theory, predicting a $|t_c|^2$ scaling of the current amplitude, and the BW formula in Eq. \eqref{eq:BW}, where the coupling $t_c$ only changes the width but not the height of the peak (for the symmetric case). This discrepancy is again an effect of the strong coupling limit we are probing.

A further facet of the complementarity between our zero-temperature simulations and the perturbative rate equation calculations is provided by the behavior of the tails of the peak as a function of the induced charge: rate equations result in a temperature-broadening of the conductance peak with exponentially suppressed tails, whereas our approach describes a broadening induced by the tunneling amplitude, which is instead characterized by a power law decay far from $n_g=0.5$.

\begin{figure}
  \begin{center}
    \includegraphics[width=\columnwidth]{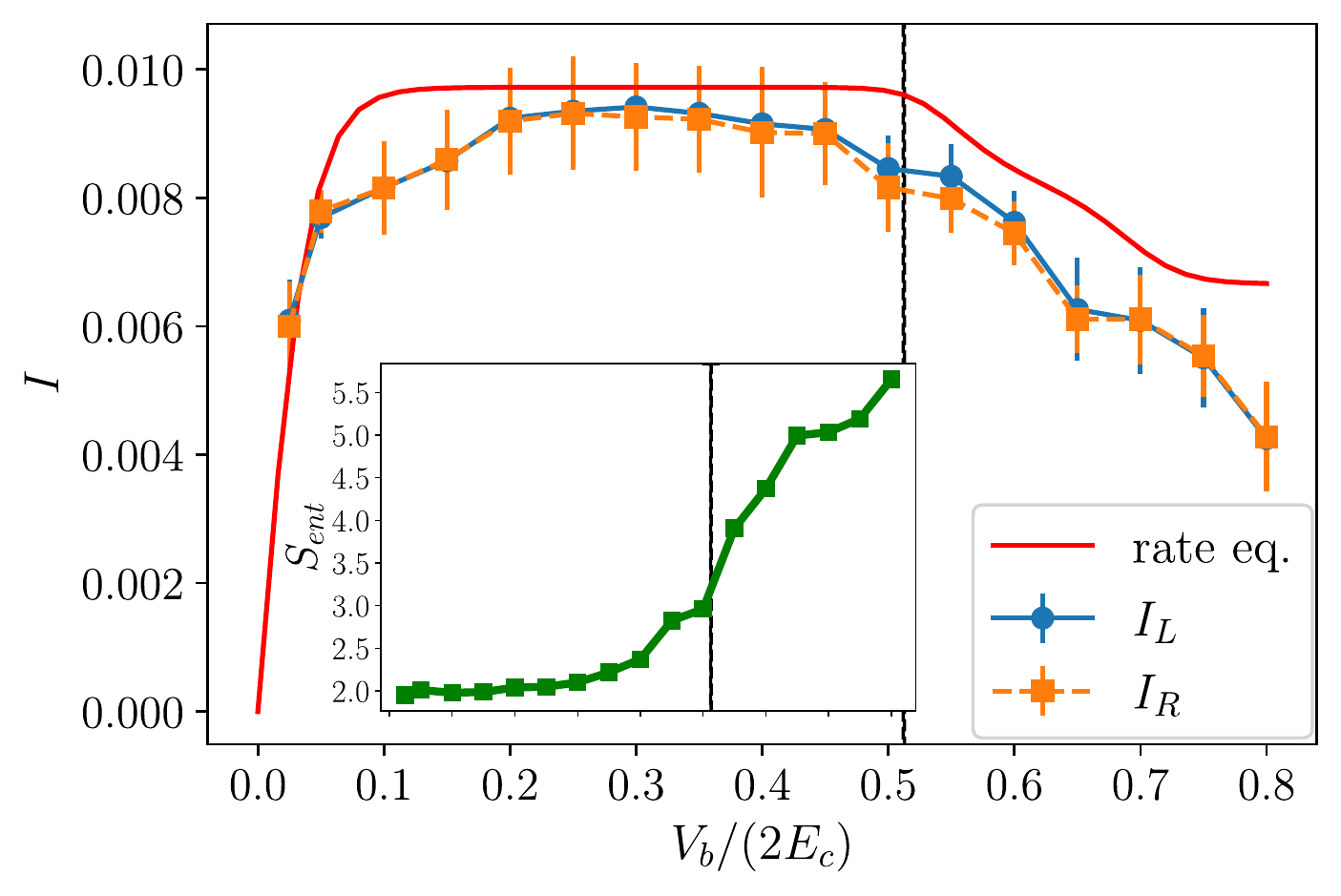} 
    \caption{Current vs voltage bias in correspondence of the induced charge resonance $n_g=0.5$ for a Kitaev chain with $\mathcal{M}=8$ sites; the vertical dashed line indicates the energy corresponding to the superconducting gap.
    The overall scale of the rate equation current (solid red line) has been arbitrary chosen to approximately match that of the TDVP algorithm.
    The inset shows the maximum entanglement entropy in the chain reached as a function of $\Vbias$. Again, the vertical dashed line indicates the position of the SC gap, where we observe a fast increase of the entanglement. For the largest two values of $\Vbias$ considered, the estimate of $S_{ent}$ is no longer reliable because the simulation saturates the maximum bond dimension allowed and the error introduced in the SVD truncation becomes larger than the chosen cutoff of $10^{-7}$.}\label{fig:IvsVb}
  \end{center}
\end{figure}

The next step towards the full characterization of the conductance in the $n_g-\Vbias$ plane is the analysis of the current with respect to the voltage bias. 
As mentioned above, here we restrict our simulations to $\mathcal{M}=8$ sites in the scatter to reduce the entanglement growth when $\Vbias$ becomes larger than the superconducting gap.
Our results are reported in Fig.~\ref{fig:IvsVb}, where we plot the current on both edges of the Kitaev chain as a function of the voltage difference between the leads, in correspondence to the resonance $n_g=0.5$.
Since we expect single-electron processes to be dominant at the charge degeneracy point, we compare our data with the prediction of first order rate equations, which indeed gives a qualitative agreement.
However, it must be noticed that both the overall amplitude of the current obtained with rate equations, and the temperature used in the corresponding Fermi factors have been arbitrary set to approximately match the profile given by the MPS data (see Appendix \ref{app:rates} for more detail).

As expected, the current increases sharply at small values of $\Vbias$ (the zero-bias peak) and then saturates in the region where transport is mostly mediated by the Majorana edge modes while the voltage is too small to excite states above the SC gap, indicated by the vertical dashed line.
When $\Vbias > \Delta_{SC}\sim 0.5 E_c$, the current decreases because transport across excited states is less efficient with respect to the Majorana modes, due to the reduced projection on the device edges.
From the point of view the differential conductance, this appears as a region with negative $G$, as observed in several experiments on SC devices with subgap states~\cite{Higginbotham_NatPhys2015,Vekris_DNW2021}.

An important effect of the excited states is the abrupt increase of the entanglement entropy at the central bond when $\Vbias$ exceeds the SC gap, as reported in the inset of Fig.~\ref{fig:IvsVb}.
Indeed when the voltage is large enough for multiple quasiparticle states to become populated, the entanglement grows accordingly.
When the bias is smaller than $\sim 0.7 E_c$, the simulations are accurate (truncation error below $10^{-7}$).
When the bias becomes large, $\Vbias/2 \gtrsim 0.7 E_c$, the system progressively saturates the maximum entanglement allowed by our simulations, which is set by the maximal bond dimension $\chi_{\rm max}=2500$, and the errors of our simulations are no longer under control.

This constitutes the principal limitation of our method:
When a continuous spectrum or a large number of excited states are within the bias energy window, thus having significant contribution to the transport of electrons, the entanglement will grow rapidly with time, hence requiring more computational resources for the accurate simulation of the dynamics, in line with general limitations of tensor network methods for studying quantum quenches.
There are several strategies to fine tune the simulations parameter and mitigate these limitations, such as choosing optimal time step durations and adjusting the localization length $\xi$ in the leads.
The best choice depends, however, on the physical parameters of the model and needs to be set accordingly.

\begin{figure}
  \begin{center}
    \includegraphics[width=\columnwidth]{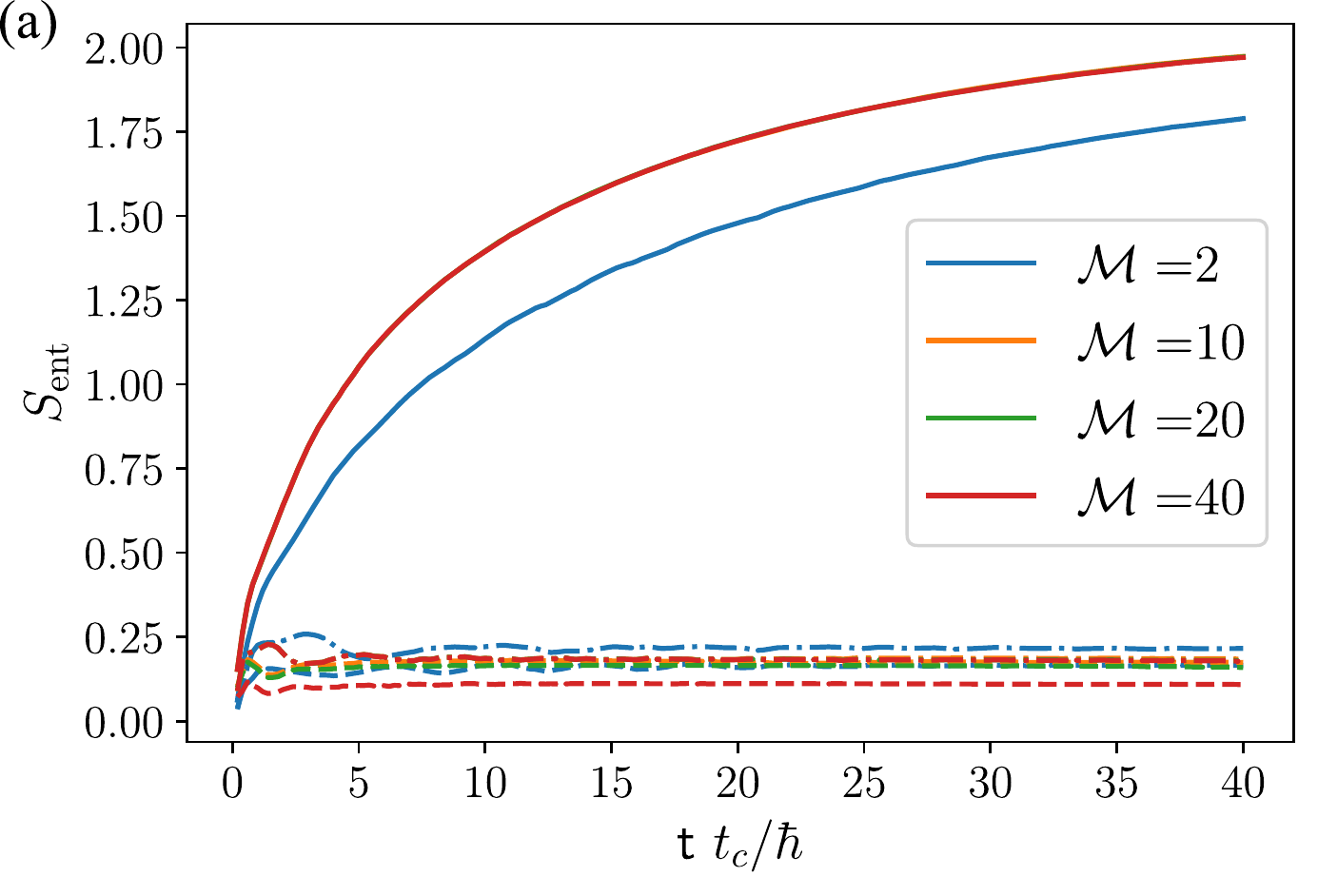}
    \includegraphics[width=\columnwidth]{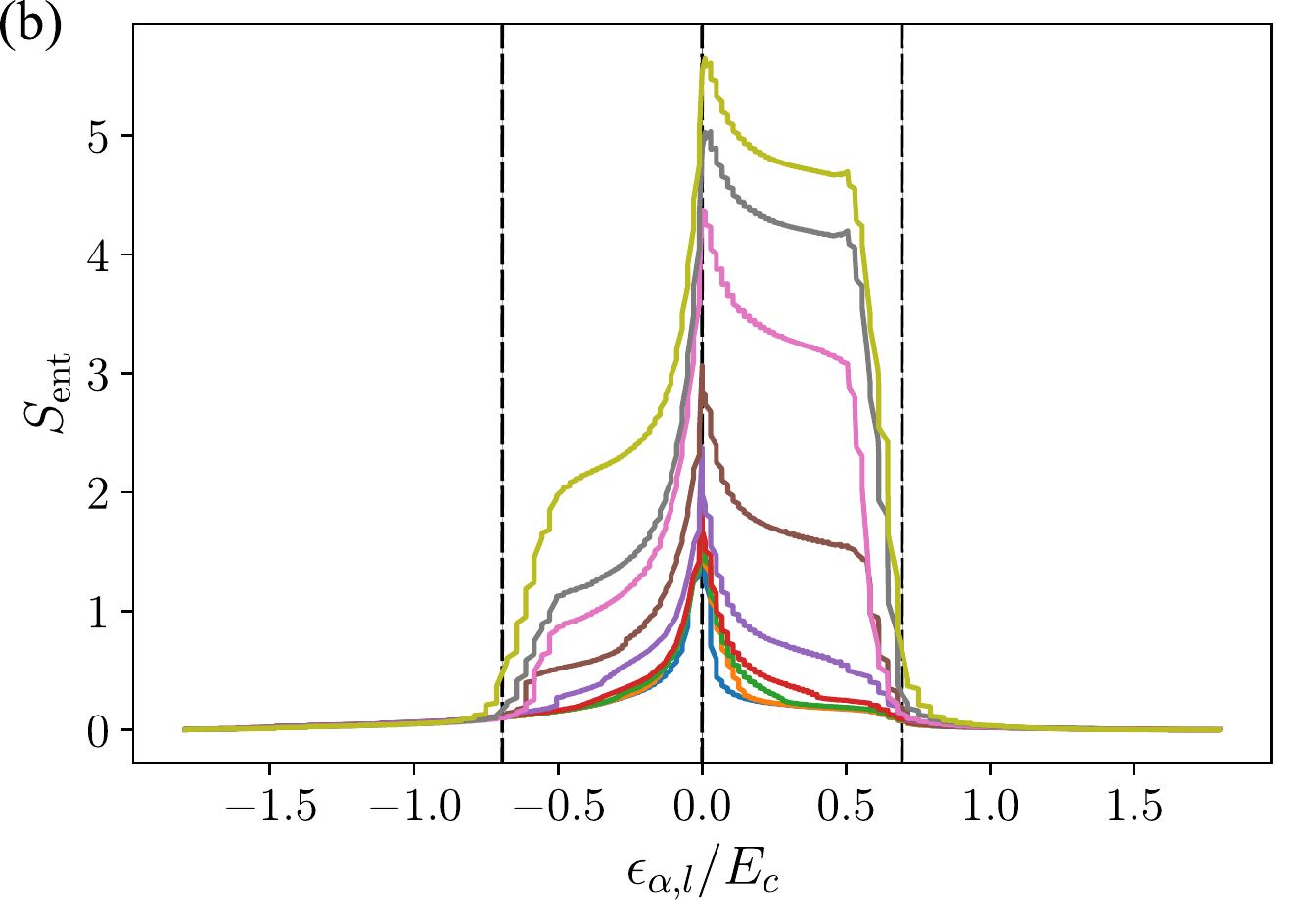}
    \caption{(a) increase of $S_{ent}$ in time for cuts in three different positions: auxiliary charge site (solid lines), right left lead bulk (dashed lines), and right lead bulk (dot-dashed lines). The color encodes the size of the scatterer; Notice that the curves corresponding to the entropy at the charge site (solid curves) are perfectly superimposed for $\mathcal{M}>2$, indicating that $S_{\rm ent}$ is almost independent from $\mathcal{M}$  as long as transport is dominated by a single channel. The voltage bias is $V_b=0.1E_c$.
    (b) entropy profile of $S_{\rm ent}$ at the end of the time evolution, for different values of the $V_b \in [0.1E_c, 1.6E_c]$. Larger entropy is associated to larger biases. The vertical dashed lines indicate the energy range of the possible manybody transitions of the scatterer, where the total charge changes by 1. This interval is centered around 0 because the data are taken at the charge degeneracy point $n_g=0.5$.
The other parameters in the Kitaev chains are $\Delta=t_s=0.3 E_c$, $\mu_s =-0.1E_c$, and $t_c=0.2E_c$.}\label{fig:entanglement}
  \end{center}
\end{figure}

The entanglement entropy at each link is therefore a useful quantity to monitor because it determines whether the simulation succeeds or not: if the entanglement entropy grows too much, so does the bond dimension $\chi$ needed to describe faithfully the quench dynamics.
$\chi$, in turn, is constrained by the memory and computational time allocated for the calculation.
On a practical level, we need to ensure that the maximum bond dimension $\chi_{\rm max}$ allowed is large enough to observe the emergence of a stationary value for the current. 

Figure \ref{fig:entanglement} (a) shows the increase in time of the entanglement entropy $S_{ent}$ corresponding to partitions of the tensor network at three different positions and for several lengths of the scatterer (color-coded).
Solid lines correspond to a cut at the position of the charge site, which lies at the center of the MPS. Dashed and dot-dashed lines instead correspond to cuts in the middle of the leads bandwidth, at negative and positive energies respectively.
The simulations are performed at the charge degeneracy point $n_g=0.5$ with a small bias $\Vbias = 0.1E_c$.
Notice that the entanglement entropy at the auxiliary charge site position, where the entropy is the largest, is almost independent from the length of the scattering region $\mathcal{M}$. Hence, in the regime where a single state mediates transport, the efficiency and speed of the simulations depend very weakly on the number of single-particle states in the SC device, allowing for studying relatively large systems.

In Fig.~\ref{fig:entanglement}(b) we plot the entanglement entropy profile as a function of the energy of the MPS sites, at the end of the time evolution.
Different curves refer to different values of $\Vbias$, corresponding to the data presented in Fig.~\ref{fig:IvsVb}, where larger voltage biases induce larger entanglement and also a wider region of the MPS where $S_{\rm ent}$ grows. 
Notice, however, that the entanglement is mostly restricted to energies limited by the bandwidth of the scattering device, indicated by the vertical dashed lines.

These data make clear the advantage deriving from our basis choice: the entanglement grows logarithmically in time and is concentrated in an energy window limited by the minimum between $\Vbias$ and the bandwidth of the scatterer. 
The ``bulk'' of the leads, meaning states far in energy from the Fermi level, almost remains in its initial product state, as seen also from Fig.~\ref{fig:occupation_N2}.
The logarithmic growth of the entanglement entropy implies that the bond dimension grows linearly in time, making the simulation efficient, since the resources required increase at most as a power law of the system size and the total evolution time.
However, the situation is not always so favorable: in general, the larger the current, the faster the entanglement and bond dimension grow.

\section{Conclusions}\label{sec:conclusion}
In this paper, we illustrated an efficient method to simulate transport phenomena in Coulomb blockaded one-dimensional superconducting systems, which encompass several fundamental building blocks for the realization of both SC qubits and many proposed platform for topologically protected qubits.
We extract their differential conductance from the quasi-steady state current arising after a quantum quench in which we bring the system out of equilibrium by imposing a finite voltage bias between two leads connected to it. 

Our method allows for exploring the system behavior in the strong coupling regime between the leads and the interacting SC device and for describing results beyond perturbative approaches (for instance, standard rate equation approaches). It is therefore suited for the study of non-perturbative phenomena such as Kondo or topological Kondo effects.

We simulate the real-time system dynamics within the MPS framework, where each site represent a single-particle energy eigenstate and the charging energy is encoded in an auxiliary site describing the total charge of the scattering region. In this basis, the system entanglement remains localized \cite{Rams2020} an we are able to compute efficiently the time evolution for long times.

In this article, we focused on simple p-wave superconducting spinless models and non-interacting leads; in particular, we analyzed two physical examples: a superconducting dot with two quasiparticle states and a blockaded Kitaev chain.  Concerning the former, our method reproduces the predicted sequential tunneling and cotunneling signatures and accounts for the interference between different coherent transport processes. Concerning the latter, our results show that in the strong coupling limit the quantization of the zero-bias differential conductance peak associated with resonant electron tunneling mediated by Majorana modes with a symmetric coupling to the leads is lost. Higher energy states suppress indeed the conductance peak while broadening its shape.

Our method can be easily extended to take into account more leads, additional interactions and spinful systems. A weak Josephson coupling between the superconducting scatterer and  grounded superconductors can also be efficiently described, as well as interacting or more complex leads. Furthermore, several recent works applied MPS methods to describe the time evolution after a quantum quench and the transport in interacting systems with leads of infinite length \cite{Zauner_2015,PhysRevB.99.121103,wu2020,PhysRevB.104.235142}. Such techniques can be integrated in our approach and can provide a method, alternative to the Wilson chain description, to mitigate finite size effects.

Our results to estimate the system conductance can also be integrated with recently developed techniques to simulate a dissipative time evolution of the system based on suitable Lindblad operators \cite{Wojtowicz2020,Brenes2020,Lotem2020}. Indeed, our simulations focused on the unitary evolution of closed systems. However, the description we have chosen for the leads degrees of freedom can be extended to account for the coupling with external electronic reservoirs, as well as other thermodynamic baths imposing a dissipative dynamics to the system. Furthermore, the introduction of suitably tailored dissipative terms that decrease the weight of non-local operators in the MPS time evolution has been proven beneficial to reduce the increase of the system entanglement in diffusive systems \cite{Pollmann2020}; such technique could be implemented also in our TDVP evolution, thus allowing for longer simulation times.

Adding the possibility of simulating the quench dynamics at finite temperature is also a necessary step to observe some of the scaling properties associated to topological \cite{Beri2012,Beri2013,Altland2013,Altland2014,Zazunov2014,Beri2017,Michaeli2017,Buccheri_TKE2020,Vayrynen_PRR2020} or charge Kondo effects~\cite{Papaj2019,Giuliano2020,han2021fractional}.
A possible route to include a finite temperature within the tensor network framework, is to promote the  system wavefunction to a density matrix, encoded in a matrix product operator (MPO), which can describe both a pure and an open mixed state~\cite{Prosen_JSTAT2009,Benenti_PRB2009}. 
Alternatively, one can keep a pure state description of the system, thus maintaining its MPS structure, by exploiting a {\em thermofield} transformation~\cite{deVega_PRA2015,Weichselbaum2018,Kohn_2022}, where the thermal distribution in the leads is reached by tracing out a set of auxiliary sites.

\section*{acknowledgements}
We warmly thank K. Flensberg, M. Leijnse, J. Paaske, R. Seoane Souto, and Mingru Yang for useful discussions. This project is supported by the Villum Foundation (Research Grant No. 25310) and received funding from the European Union’s Horizon 2020 research and innovation program under the Marie Sklodowska-Curie grant agreement No. 847523 “INTERACTIONS.” C.-M.C. acknowledges the support by the Ministry of Science and Technology (MOST) under Grant No. 111-2112-M-110-006-MY3, and by the
Yushan Young Scholar Program under the Ministry of Education (MOE) in Taiwan.

\appendix

\section{Details on the MPS construction} \label{app:MPS}

\begin{figure}
\includegraphics[width=\columnwidth]{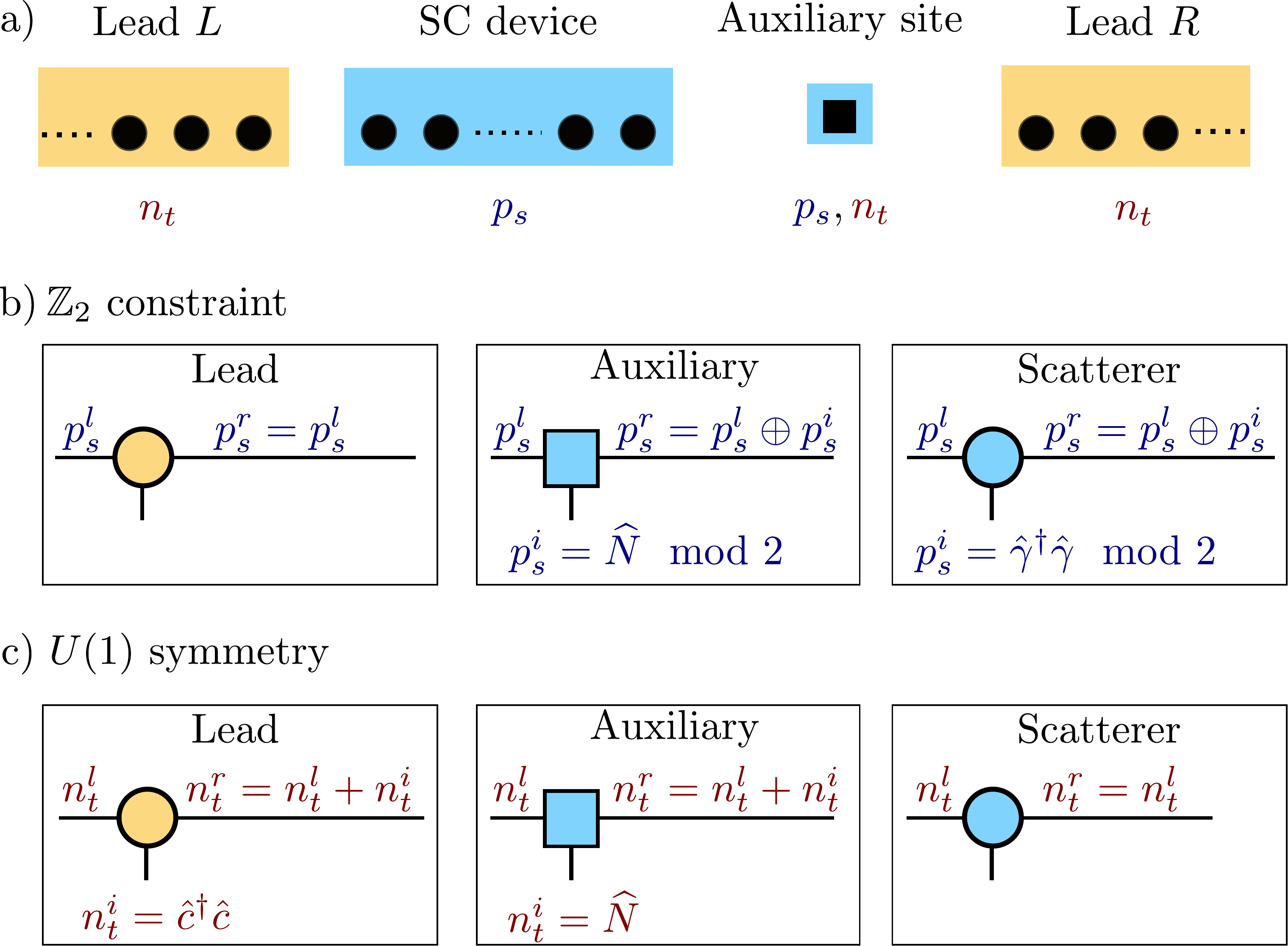}
\caption{a) The quantum numbers $p_s$ and $n_t$ refer to the fermionic parity of the scatterer and the total particle number respectively. The scheme indicates for which kind of sites in the MPS these quantum numbers are defined. b) The $\mathbb{Z}_2$ parity constraint \eqref{constr} is imposed by building tensors fulfilling the depicted rules. The lead tensors connect virtual states with the same parity. The auxiliary and scatterer sites, instead, modify the fermionic parity of the virtual states accordingly to the physical degree of freedom, such that they are $\mathbb{Z}_2$ invariant. c) The global $U(1)$ gauge invariance is enforced by assigning to each virtual link in the MPS chain a virtual charge $n_t$. For the auxiliary and lead sites, the tensors increase the virtual charge based on the charge of the physical state $n_t^i$. The scatterer sites, instead, leave the virtual charge unaltered.} 
\label{fig:MPS}
\end{figure}

Our MPS construction, represented in Fig. \ref{fig:sketch}, exactly implements the parity constraint given by Eqs. \eqref{parityc} and \eqref{constr} and the $U(1)$ global symmetry associated to the conservation of the charge in Eq. \eqref{Ntot}. These are two independent symmetry conditions which the MPS construction must fulfill, and we emphasize that the first involve only the physical degrees of freedom of the scatterer and the auxiliary state, whereas the second involves only the physical degrees of freedom of the leads and the auxiliary site.
The tensors entering the description of the system must accordingly fulfill the $\mathbb{Z}_2$ and $U(1)$ charge conservation rules represented in Fig. \ref{fig:MPS}, which involve both the physical and virtual indices (see \cite{Singh2010PRA,Zohar2016_NJP,Silvi_SCIPOST2019} for general overviews of symmetries in tensor networks). These requirements are set by assigning two independent quantum numbers, which we label by $p_s$ and $n_t$ and correspond to the scatterer fermionic parity and the total particle number, to all the  states in each (virtual or physical) bond of the MPS.

Fig. \ref{fig:MPS}(a) summarizes which kind of physical sites contribute to the quantum numbers $p_s$ and $n_t$ of the MPS tensors.
The quantum number $p_s = 0,1$ concerns the $\mathbb{Z}_2$ fermionic parity of the scatterer degrees of freedom. In particular, $p_s$ on the bond $l$ refers to the fermionic parity accumulated in the scatterer physical sites (including the auxiliary charge site) on the MPS sites before the bond $l$. In Fig. \ref{fig:MPS}(b), $p_s^{l/r}$ labels left and right virtual states of the tensors respectively; $p_s^i$ refers instead to the fermionic parity of the physical sites of the scatterer. In order for the constraint \eqref{constr} to be satisfied, the lead tensors must connect left and right virtual states with the same parity $p_s$, irrespective of the lead physical degree of freedom [since Eq. \eqref{constr} does not involve them]; the auxiliary and scatterer site tensors, instead, must connect virtual left and right states with different parity whenever the physical fermionic parity of the site is odd, and leave the virtual parity $p_s$ invariant when the physical parity is even. This corresponds to the conservation of the fermionic parity $p_s$ depicted in Fig. \ref{fig:MPS}(b). The constraint \eqref{constr} is enforced by a suitable choice of the tensor boundary conditions.

Concerning the global $U(1)$ charge conservation, instead, the physical degrees of freedom involved are the ones of the leads and the auxiliary site. Such charge conservation is imposed by assigning a further charge quantum number $n_t$ to all \sout{virtual} states in the MPS bonds. In this case, the lead and auxiliary site tensors must suitably increase this virtual charge from the left to the right tensor link based on the charge of the physical site. The scatterer site tensors, instead, connect virtual states with the same virtual charge. See Fig. \ref{fig:MPS}(c). 

We emphasize that $p_s$ must not be confused with the parity of $n_t$. Clearly, since Eq. \eqref{Ntot} sets a $U(1)$ symmetry of the system, it also enforces an additional parity conservation for the lead and auxiliary site degree of freedom, which corresponds the $\mathbb{Z}_2$ symmetry that characterizes the system when introducing the additional Josephson coupling $E_J$ in Eq. \eqref{SCbox}. In this case, the constraints in Fig. \ref{fig:MPS}(c) are relaxed to analogous relations involving only the new parity $q_t = n_t \mod 2$.

\section{Rate equation approach}\label{app:rates}
To compare our numerical results with an analytical approach, we use a standard rate equation approach, where we consider a classical probability distribution for the occupation numbers of the quasiparticle (QP) states in the scatterer and transition rates given by Fermi golden rule.

First, let us rewrite the Hamiltonians of the quantum dot:
\begin{equation}
\Hsys = \sum_{n=1}^N \epsilon_n \opgammadag{n} \opgamma{n} + \mathcal{E}_0 + E_c(N-n_g)^2 \ ,
\end{equation}
where $\mathcal{E}_0$ is the energy of the BdG vacuum and $\opgamma{n}$ is the destruction operator of the quasiparticle state with positive energy $\epsilon_n$.

Regarding the leads, we make a wide-band approximation, with a linear dispersion and a constant density of states
\begin{equation}
\Hleads = \sum_{\alpha=R,L} \sum_k \xi_{\alpha,k}\opcdag{\alpha,k}\opc{\alpha,k} \ , 
\end{equation}
where we assume that the eigenstates are plane waves with momentum quantization $k=\frac{2\pi}{L} j$, with $j$ integer. To link this approximation with the original tight binding description we fix $\xi_{\alpha,k} = \hbar v^F_{\alpha} k-\mu_\alpha = 2t_0^\alpha  ka-\mu_\alpha$, being $2t_0^\alpha a/\hbar$ the lead Fermi velocity at half filling and $a$ the lattice spacing.

Finally, the tunneling Hamiltonian between lead $\alpha$ and the SC device reads
\begin{equation}
\Htunn^\alpha = -t_{c,\alpha} \sum_{k,n} \left[ \left(\opgammadag{n} u^*_n (x_\alpha) +\opgamma{n} v_n(x_\alpha) \right)\opc{k} \phi_{\alpha,k}  +\hc \right].
\end{equation}
Here, $u^*_n(x_\alpha)$ and $v_n(x_\alpha)$ are the particle and hole weights, respectively, of the $n-$th eigenstate on the first $(x_L=1)$ or last $(x_R=N)$ site of the chain, while $\phi_{\alpha,k}$ are the plane waves states of the leads.

Since we are interested in a strong Coulomb blockaded regime, we restrict the scatterer Hilbert space considering only states with total charge $N=0,2$, 
with no QP excitation, and $N=1$, with a single QP state occupied. 
We will denote with $P_0$, $P_2$ and $P_n$ the populations of such states.

Sequential tunneling events connect states with different quasiparticle occupation. hence we need to compute the four transition rates $\Gamma^\alpha_{0n}$, $\Gamma^\alpha_{n0}$, $\Gamma^\alpha_{2n}$, $\Gamma^\alpha_{n2}$ for both contacts with the external leads.
Following Fermi golden rules, we can compute these rates as 
\begin{equation}
\Gamma_{if} = \frac{2\pi}{\hbar} |\bra{f}\Htunn\ket{i}|^2 w_i \delta(E_i-E_f)\ ,
\end{equation}
where $\ket{i}$ and $\ket{f}$ are the initial and final states and $w_i$ the thermal weight of the initial states.
Standard calculations lead to
\begin{eqnarray}\label{eq:rates}
\Gamma^\alpha_{0n} &= \frac{\pi}{h} \frac{t_{c,\alpha}^2}{t_0}|u_n(x_\alpha)|^2 f(\Delta E_{10} + \epsilon_n -\mu_\alpha) \ , \\
\Gamma^\alpha_{n0} &= \frac{\pi}{h} \frac{t_{c,\alpha}^2}{t_0}|u_n(x_\alpha)|^2 \left[1-f(\Delta E_{10} + \epsilon_n -\mu_\alpha)\right] \nonumber \ ,\\
\Gamma^\alpha_{2n} &= \frac{\pi}{h} \frac{t_{c,\alpha}^2}{t_0}|v_n(x_\alpha)|^2 \left[1-f(\Delta E_{21} - \epsilon_n -\mu_\alpha)\right] \nonumber \ , \\
\Gamma^\alpha_{n2} &= \frac{\pi}{h} \frac{t_{c,\alpha}^2}{t_0}|v_n(x_\alpha)|^2 f(\Delta E_{21} - \epsilon_n -\mu_\alpha) \nonumber \ .
\end{eqnarray}
$\Delta E_{N,N'}$ is the charging energy difference between two states with different total charge and $f(\cdot )$ is the Fermi distribution.
Once the rates are known, we can derive the nonequilibrium steady state by looking for the kernel of the transition matrix obtained from the rates in Eq.~\eqref{eq:rates}.

An important thing to notice is the asymmetry between the transition $0 \leftrightarrow n $ and $ 2\leftrightarrow n$. Indeed, in the first, a particle is directly transferred from the leads to a quasiparticle state, or vice-versa, while, in the second, the process involves the destruction (creation) of a Cooper pair.
If the particle and hole weights happen to be very different, for instance in a Kitaev chain in the topologically trivial phase, this asymmetry is reflected in the Coulomb diamonds.
Indeed, the current associated to direct tunneling of an electron in a quasiparticle state would be much larger than the current associated to process involving a destruction or creation of a Cooper pair.

Once the rates are computed, the stationary  probability distribution for the states of the SC island $P^{\rm eq}$ are found by solving the linear system of equations
\begin{equation}\label{eq:Peq}
\left\{
\begin{split}
&\sum_n \Gamma_{n0} P^{\rm eq}_n - \Gamma_{0n}P^{\rm eq}_0  = 0 \ ,\\
&\Gamma_{0n}P^{\rm eq}_0 + \Gamma_{2n}P^{\rm eq}_2 - (\Gamma_{n0}+\Gamma_{n2}) P^{\rm eq}_n  = 0 \ ,\\
&\sum_n \Gamma_{n2} P^{\rm eq}_n - \Gamma_{2n}P^{\rm eq}_2  = 0 \ ,
\end{split}
\right.
\end{equation}
where $\Gamma_{n0} = \Gamma_{n0}^L+\Gamma_{n0}^R$ and similarly for the the other rates.
Notice that Eq.~\eqref{eq:Peq} only considers states with one or no QP state occupied, which is valid only at small voltage bias.
In order to obtain a more accurate description and take into account also transitions between excited states, situations in which tow or more quasiparticle are present should also be considered. This has been done, for instance, to obtain all the main sequential tunneling resonances appearing as diagonal lines in Fig. \ref{fig:SCdot} (b).
Such extensions of the considered many-states involved in the transport, however, rapidly increase the dimension and the complexity of the transfer matrix describing the incoherent evolution of the scatterer, making it intractable for systems with more than a few sites, despite they do not complicate the calculation of the rates themselves.

Combining the definition of the transition rates in Eq.~\eqref{eq:rates} and the associated probability distribution from Eq.~\eqref{eq:Peq}, the sequential tunneling contribution to the current from lead $\alpha$ to the device reads
\begin{equation}\label{eq:I_seq}
I_\alpha = e \sum_n (\Gamma^\alpha_{n2} -\Gamma^\alpha_{n0}) P^{\rm eq}_n +\Gamma^\alpha_{0n}P^{\rm eq}_0 - \Gamma^\alpha_{2n}P^{\rm eq}_2 \ ,
\end{equation}
where we adopted the convention that an ingoing particle current is positive. If the system has only two terminals, $I_L = -I_R$ using this convention.
In the main text, we make a different choice for the current sign: a positive (particle) current flows from the left lead to the device and then to the right lead, while a negative current flows in the opposite direction.
An important thing to notice about eqs.~\eqref{eq:Peq} and \eqref{eq:I_seq} is that $P^{\rm eq}$ is independent from the tunneling strength $t_c$ between the leads and the SC island while the current  inherits a global factor $|t_c|^2$ from the rates $\Gamma$, which sets the overall scaling of $I_\alpha$ with $t_c$.
Moreover both $P^{\rm eq}$ and $I_\alpha$ require a finite temperature $T$ to be well defined and to avoid discontinuities in the rates and in the stationary probability distribution due to sharp jumps in the Fermi factors at $T=0$.
Hence, it is impossible to compare quantitatively the result of the perturbative rate equations and our exact solution of the unitary dynamics using tensor networks at zero temperature.
The former requires $T>0$ and only describes a $|t_c|^2$ scaling of the current. Moreover the conductance resonances are broadened only by the temperature and not by the finite coupling between the leads and the scatterer.
Our MPS approach, instead, works at $T=0$ and predicts a coupling-induced broadening of the conductance peaks as well as a less trivial scaling of the current amplitude with $t_c$.
Thus, when comparing qualitatively the two approaches, it is necessary to choose an appropriate scaling factor if one wishes the current signals to have similar amplitudes, as we did in Fig. \ref{fig:IvsVb}.

\bibliography{MajoranaBiblio}
\end{document}